\documentclass[epjST]{svjour}
\usepackage{geometry}
\usepackage[T1]{fontenc}
\usepackage{lmodern}

\usepackage[utf8]{inputenc}
\usepackage{amsmath, amssymb}
\usepackage{physics}
\usepackage{graphicx}
\usepackage{caption}
\usepackage{subcaption}
\usepackage{float}
\usepackage{todonotes}
\usepackage[comma,numbers,sort&compress]{natbib}
\usepackage{enumitem}
\usepackage{hyperref}
\usepackage{ulem}
\usepackage{placeins}

\graphicspath{{./figures/}}
\newcommand{\scale}{0.5}
\newcommand{\Lscale}{0.8}
\newcommand{\pa}{p_{p\rightarrow a}}
\newcommand{\pas}{\tilde p_{p\rightarrow a}}
\newcommand{\pp}{p_{a\rightarrow p}}

\newcommand{\p}{p_{o\rightarrow o'}}
\newcommand{\ps}{\tilde p_{o\rightarrow o'}}

\colorlet{darkgreen}{olive!140!}
\definecolor{cadmium}{rgb}{0.0, 0.42, 0.24}
\definecolor{darkorange}{rgb}{1.0, 0.55, 0.0}
\definecolor{deepcarrotorange}{rgb}{0.91, 0.41, 0.17}

\newcommand{\jobst}[1]{\textcolor{black}{#1}}
\newcommand{\jv}[1]{\textcolor{black}{#1}}
\newcommand{\jona}[1]{\textcolor{black}{#1}}
\newcommand{\jakob}[1]{\textcolor{black}{#1}}
\newcommand{\niklas}[1]{\textcolor{black}{#1}}
\newcommand{\marc}[1]{\textcolor{black}{#1}}

\begin{document}
\title{\jona{Dose-response functions and surrogate models for exploring social contagion in the Copenhagen Networks Study}}

\author{\jona{Jonathan F. Donges\inst{1,2,*}\fnmsep\thanks{\email{donges@pik-potsdam.de}, ORCID: 0000-0001-5233-7703} \and Jakob H. Lochner\inst{1,3,*} 
\and Niklas H. Kitzmann\inst{1,4}\fnmsep\thanks{ORCID: 0000-0003-4234-6336}
\and Jobst Heitzig\inst{1} 
\and Sune Lehmann\inst{5,6} 
\and Marc Wiedermann\inst{1,7,8} 
\and J\"urgen Vollmer\inst{3}\fnmsep\thanks{ORCID: 0000-0002-8135-1544}
}}
\institute{Earth System Analysis \& Complexity Science, Potsdam Institute for Climate Impact Research, Member of the Leibniz Association, 14473 Potsdam, Germany, EU \and Stockholm Resilience Centre, Stockholm University, 10691 Stockholm, Sweden, EU \and Institute for Theoretical Physics, University of Leipzig, 04103 Leipzig, Germany, EU \and Institute for Physics and Astronomy, University of Potsdam, 14476 Potsdam, Germany, EU \and Department of Applied Mathematics and Computer Science, Technical University of Denmark, Lyngby, Denmark, EU \and Department of Sociology, University of Copenhagen, Copenhagen, Denmark, EU \and \marc{Robert Koch Institute, Berlin, Germany, EU \and Institute for Theoretical Biology, Humboldt University of Berlin, Berlin, Germany, EU}\\ * The first two authors share the lead authorship.}
\abstract{
\jona{Spreading dynamics and complex contagion processes on networks are important mechanisms underlying the emergence of critical transitions, tipping points and other nonlinear phenomena in complex human and natural systems.}
Increasing amounts of temporal network data are now becoming available to study such spreading processes of behaviours, opinions, ideas, diseases and innovations to test hypotheses regarding their specific properties.
To this end, we here present a methodology based on dose-response functions and hypothesis testing \jona{using surrogate data models that randomise most aspects of the empirical data while conserving certain structures relevant to contagion, group or homophily dynamics.}
We demonstrate this methodology for synthetic temporal network data \jona{of spreading processes} generated by the adaptive voter model.
Furthermore, we apply it to empirical temporal network data from the Copenhagen Networks Study. 
This data set provides a physically-close-contact network between \jona{several hundreds of} university students participating in the study over the course of three months. %\jona{with daily time resolution}. 
We study the potential spreading dynamics of the health-related behaviour ``regularly going to the fitness studio'' on this network.
\jona{Based on a hierarchy of surrogate data models, we find that our method neither provides significant evidence for an influence of a dose-response-type network spreading process in this data set, nor significant evidence for homophily.}
The empirical dynamics in exercise behaviour  
are likely better described by individual features such as the disposition towards the behaviour, and the persistence to maintain it, as well as external influences affecting the whole group, and the non-trivial network structure.
The proposed methodology is generic and promising also for applications to other \jona{temporal network data sets} and traits of interest.
} %end of abstract
\maketitle

\section{Introduction}
\label{sec:intro}
Spreading \jona{and} complex contagion processes shape the dynamics of diverse complex ecological, societal and technological systems studied in many fields of research \cite{watts2002simple,Dodds2004universalBehavior,lehmann2018complex}. 
Examples include biological infections \cite{Murray2002,Daley1999} such as the spreading of the COVID-19 pandemic \cite{maier2020effective}\niklas{; }%,
cascading failures in interdependent infrastructure systems \cite{buldyrev2010catastrophic}\niklas{; } %,
diffusion of innovations and technologies \cite{coleman1966medical,Valente1996,geels2017sociotechnical}\niklas{;} %,
\niklas{evolutionary processes \cite{capraro_mathematical_2021, turchin_war_2013}}\niklas{; } %,
social norms \cite{nyborg2016social}\niklas{,  behaviours \cite{tsvetkova_social_2014}}, and other social, political and technological innovations relevant for sustainability transition and rapid decarbonisation \cite{tabara2018positive, farmer2019sensitive, otto2020social, sharpe2021upward}\niklas{; } %,
political changes~\cite{Lohmann1994}\niklas{; } %,
or religious missionary work~\cite{Stark1996,montgomery1996diffusion}. 
These spreading processes on complex networks often give rise to nonlinear dynamics and the emergence of macroscopic phenomena, such as phase transitions and tipping points that separate qualitatively different dynamical regimes \cite{winkelmann2020social}; for example, a transition between regimes where a local infection or innovation is locally contained, and those where it spreads globally to a large part of the network \cite{watts2002simple, Dodds2004universalBehavior, Dodds2005generalizedModel, wiedermann2020network, geels2017sociotechnical}. 
Furthermore, spreading processes can interact with the underlying complex network structures, e.g.\ through the process of homophily, giving rise to complex coevolutionary feedbacks between dynamics on and structure of these networks \cite{holme2006nonequilibrium, gross2006epidemic, gross2009adaptive,wiedermann2015macroscopic}. 
Better understanding of such complex spreading processes, based on improved methods for data analysis and modelling, is highly relevant for finding robust approaches to identify, analyse, influence or govern their dynamics. 
This way, harmful impacts may be avoided, or desirable outcomes reached, e.g.\ for containing pandemic outbreaks \cite{hsiang2020effect, maier2020effective, schlosser2020covid}, preventing cascading failures in power grids \cite{buldyrev2010catastrophic,menck2014dead}, or fostering the spreading of social-cultural-technological innovations towards a rapid sustainability transformation \cite{tabara2018positive, farmer2019sensitive, otto2020social, winkelmann2020social}.

In recent years, temporal network data has become more abundantly available from social media platforms such as Facebook \cite{lewis_tastes_2008} and Twitter \cite{suh_want_2010}, or long-term health studies such as the Framingham Heart Study \cite{feinleib_framingham_1975} that have been leveraged for studying spreading and contagion processes, e.g.\ in the dynamics of obesity \cite{Christakis2007}, smoking \cite{Christakis2008}, happiness~\cite{Fowler2008}, loneliness~\cite{Cacioppo2009}, alcohol consumption~\cite{Rosenquist2010}, depression \cite{Rosenquist2011}, divorce \cite{McDermott2013}, emotional contagion \cite{kramer2014experimental} and political mobilisation \cite{bond201261}. 
So far such studies of empirical temporal network data mainly relied on standard statistical methods such as generalised linear models, generalised estimating equations or spatial autoregressive models \cite{lehmann2018complex}.
However, these methods are typically not well-equipped to deal with network dependencies \cite{ogburn2018challenges}. 
Furthermore, analogous to the problem of identifying causal associations in multivariate time series data \cite{runge2015identifying,runge2018causal}, there are challenges in extracting possible causal effects induced by contagion processes, and in separating their imprints from other mechanisms such as homophilic rewiring of network structure, common external forcing from the system's environment and other confounding effects.
After all, most studies rely on observational data and not on controlled experiments \cite{ogburn2018challenges}.

Here, we contribute to this field by developing a methodology for the analysis of complex spreading processes in  temporal network data sets based on dose response functions (DRFs) that have been used in the theoretical description of simple and complex contagion processes \cite{Dodds2004universalBehavior, Dodds2005generalizedModel}.
Among others, they have been applied to the study of behavioural contagion in animal systems such as startling cascades in fish schools \cite{sosna2019individual} and the spread of information on social media networks \cite{hodas_simple_2014}. 
Dose response functions encode a network nodes' probability of being infected with a new trait, given the level of exposure to this trait in its network neighbourhood. 
We propose an algorithm including Gaussian filtering to robustly estimate DRFs from synthetic and empirical temporal network data, including the possibility of propagating various types of uncertainties. 
In order to test for the possibility of an actual causal spreading process being involved in generating the data, and to identify confounding effects, we also develop a hierarchy of temporal network surrogate models.
\niklas{These models comprise a family of methods that rely on partial data randomisation to analyse specific features of (networked) processes without assuming particular underlying mechanisms
%\jv{They are} %making them 
\jakob{and have been proven}
highly useful in exploratory data analyses \cite{vicente_transfer_2011,casdagli_chaos_1992}.
\jv{In particular,} 
they have been used extensively to investigate temporal networks \cite{gauvin_randomized_2020, holme_temporal_2012}, including epidemic and social contagion processes \cite{genois_compensating_2015, karimi_threshold_2013}.
A conceptually related application for surrogate models is the study of time series data \cite{theiler_testing_1992, schreiber_surrogate_2000}.
Here, we combine methods from both temporal network and time series surrogate models.
This enables us}
%They enable us 
to investigate which features and structures in the data are possibly sufficient to explain the obtained dose response functions. 

We apply \jakob{our} methodology to synthetic data from the adaptive voter model as a proof-of-concept, and to empirical observational temporal network data from the Copenhagen Networks Study. 
Based on the latter we analyse the spreading dynamics of the illustrative behaviour of ``regularly going to the fitness studio'' on a physically-close-contact network between university students participating in the study over the course of three months
\jakob{with daily time resolution.}
We do not find robust evidence of a causal spreading process underlying the observed dynamics. 
This suggests that possible social contagion effects in this context are limited, and dominated by other factors or shadowed by excessive noise.
This is in agreement with findings from health behaviour psychology \cite{marcus_transtheoretical_1994}.
Hence, this first application study suggests that the proposed methodology is generic and promising for investigations of other data sets and possibly spreading traits of interest.

This paper is structured as follows: we first introduce the synthetic and empirical temporal network data sets, obtained from the adaptive voter model and the Copenhagen Network Study, respectively (Sect.~\ref{sec:data}). 
In a next step, we describe the methodology developed here for data analysis, including estimating dose response functions and generating surrogate data sets for testing hypotheses on underlying data generating processes (Sect.~\ref{sec:methods}). 
Finally, we report results obtained for the synthetic and empirical data sets (Sect.~\ref{sec:results}), discuss these findings and conclude (Sect.~\ref{sec:disc_concl}).

\section{Data}
\label{sec:data}
Here we describe the data sets used in this study to test our proposed dose-response function methodology. The data has the form of temporal networks (Sect.~\ref{sec:temporal_nets}), it includes synthetic temporal network data generated by the adaptive voter model (Sect.~\ref{sec:avm}) and empirical temporal network data from the Copenhagen Networks Study (Sect.~\ref{sec:cns}).

\begin{figure}[h]
    \centering
    \includegraphics[width=0.4\textwidth]{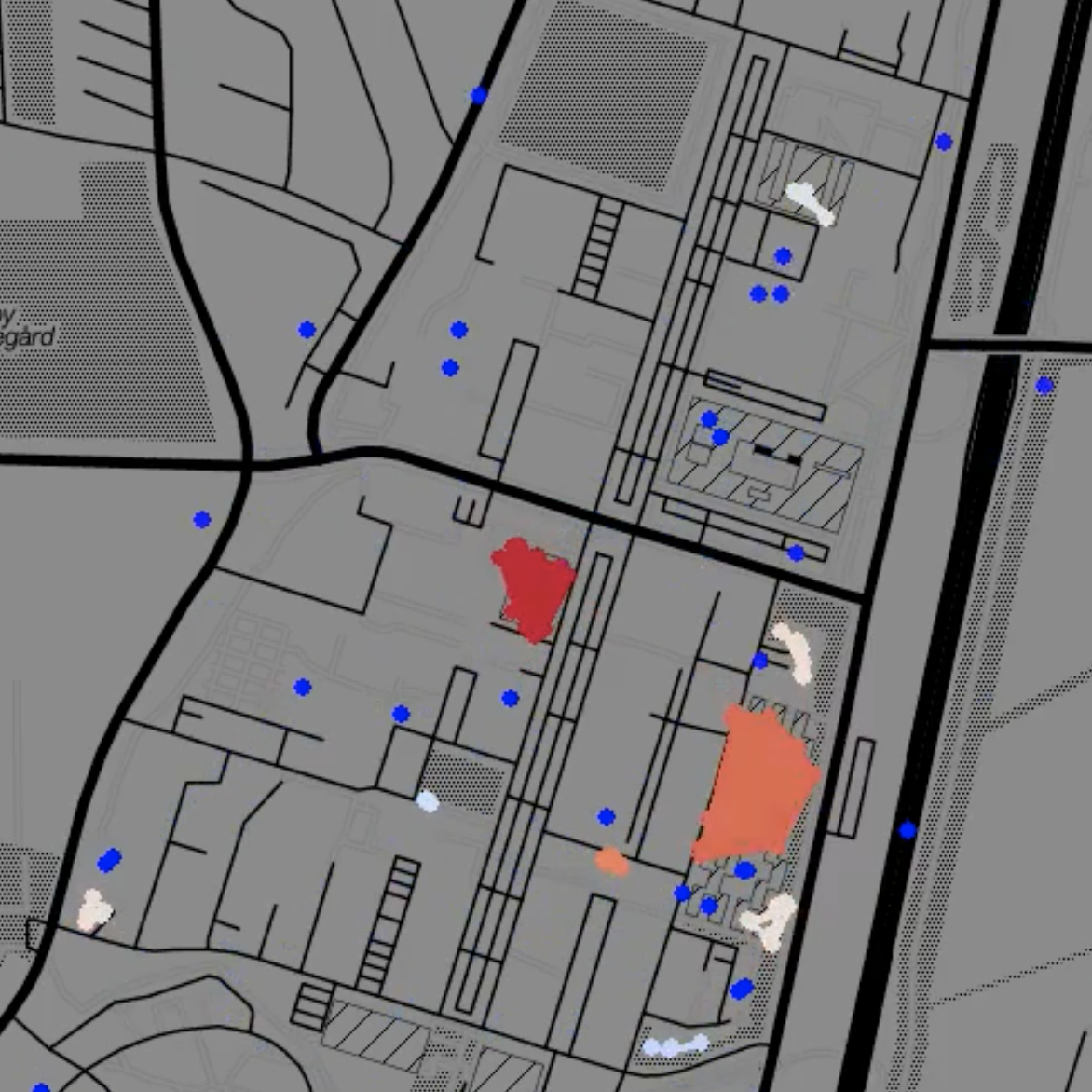}
    \includegraphics[width=0.4\textwidth]{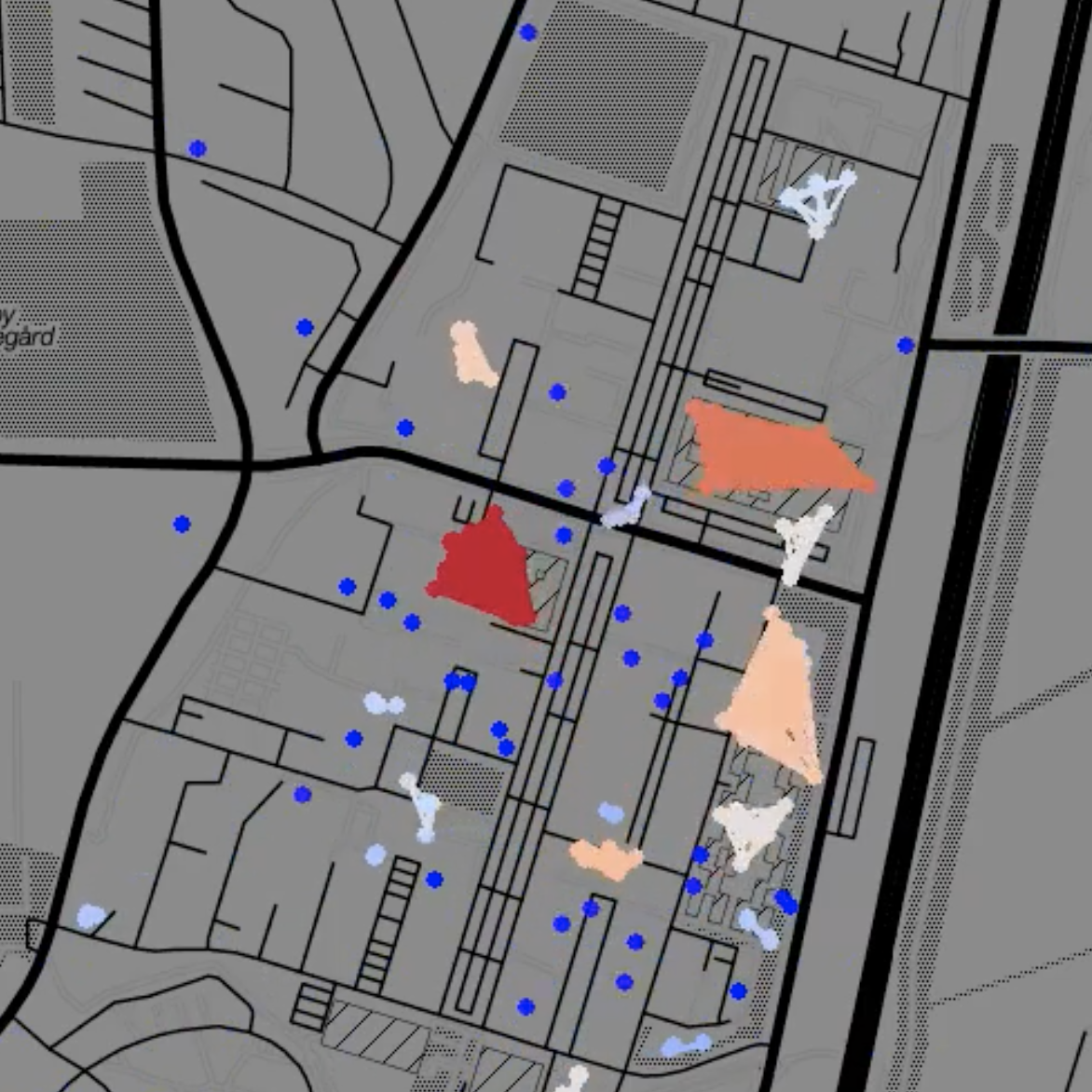}
    \includegraphics[width=0.4\textwidth]{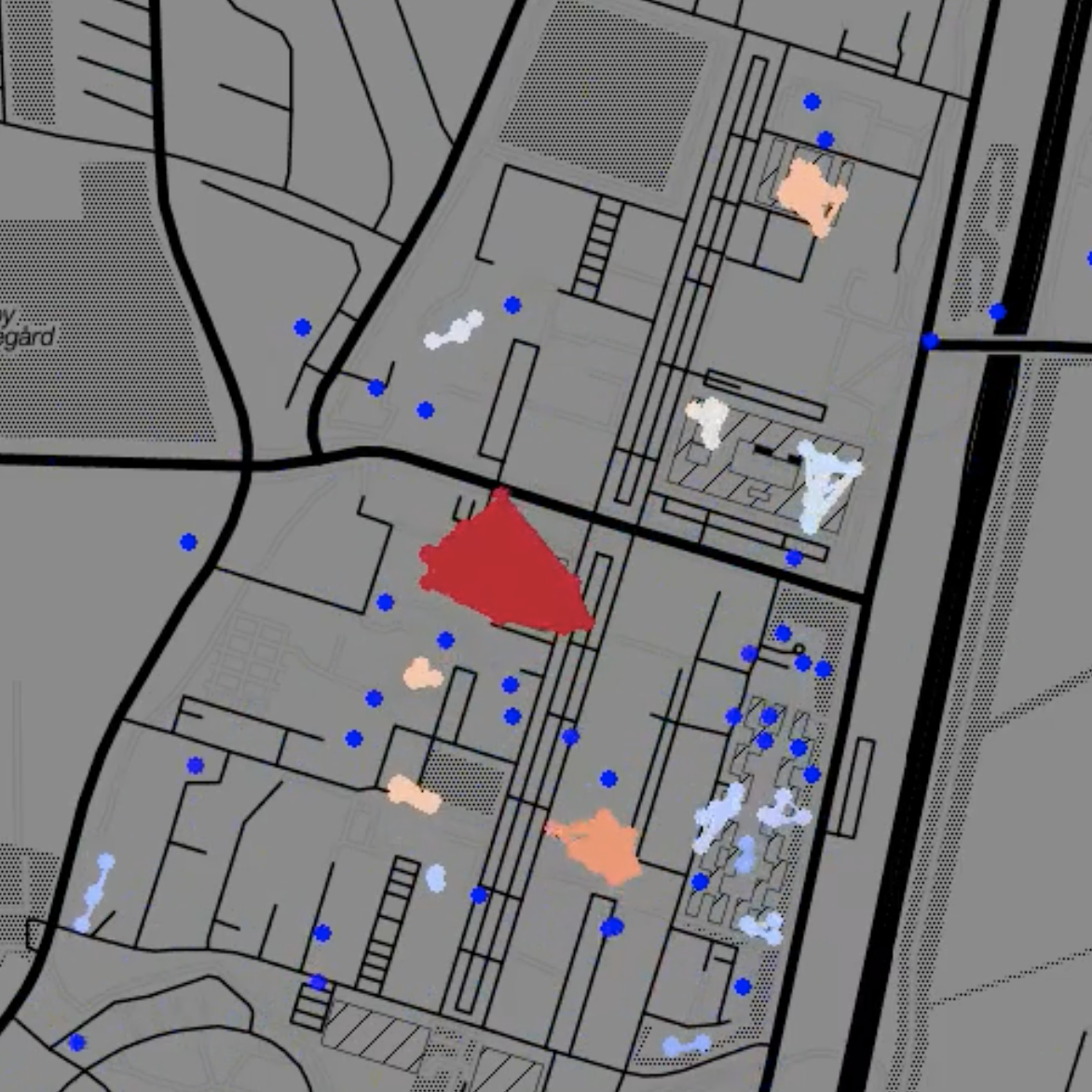}
    \includegraphics[width=0.4\textwidth]{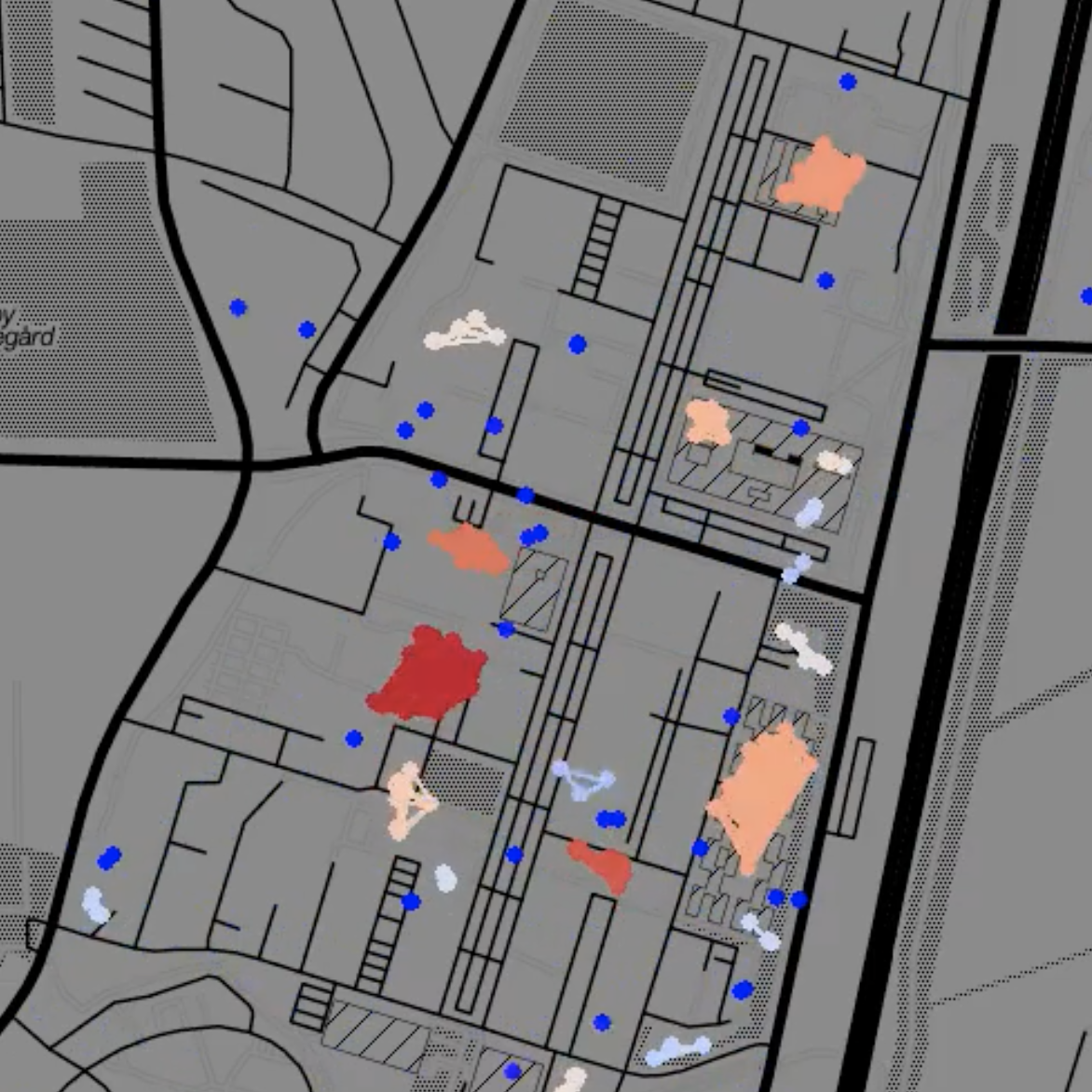}
    \caption{Temporal network snapshots throughout a typical day during the first semester of the Copenhagen Networks Study. Each dot represents an individual, colour coded according to cluster size from single nodes (dark blue) to large clusters (dark red). Node clusters evident in the snapshots correspond to students engaging in joint activities, such as lectures or eating lunch in a cafeteria.}
    \label{fig:cns_snapshots}
\end{figure}

\subsection{Temporal social networks}
\label{sec:temporal_nets}

The data sets investigated in this work are structured as temporal networks $\mathcal{G}(t)$ with a fixed number of nodes $N$ and a time-dependent set of links described by the adjacency matrix $A_{ij}(t)$, where $i,j\in\{1, \dots, N\}$ \cite{Holme2012}, sampled at discrete time steps $t$. In addition, node traits $o_i(t)$ are time-dependent as well, for example encoding \jakob{different} opinions or behaviours.

\subsection{Synthetic temporal network data: adaptive voter model}\label{sec:avm}

One prototypical model of temporal network dynamics is the adaptive voter model (AVM)~\cite{holme2006nonequilibrium} that incorporates core processes in social systems, i.e., homophily~\cite{boguna2004models} and social learning of traits~\cite{castellano2003incomplete}. As such, the AVM can be interpreted as a straightforward generalisation of the so-called voter model~\cite{Holley_ergodic_voter} to any prescribed initial social network topology and the ability of the represented individuals to deliberately change their neighbourhood structure. It thereby aims to explain the emergence of like-minded communities within a larger social network and the extent to which individuals (i) become like-minded because of shared social ties or (ii) form such social ties because they are like-minded. 

\jona{
We use an AVM to generate synthetic temporal network data that resembles the experimental data from the Copenhagen Networks Study.
This choice has several motivations:
Firstly, it matches our initial hypothesis that a quasi-symmetric social learning process underlies the spread of “active” and “passive” behaviours of individuals.
Under this hypothesis, individuals can equally imitate active or passive behaviour occurring in their network neighbourhood.
This is in contrast to standard SI(S/R)-type models~\cite{gross2008adaptive,perra2012activity}, where only one trait spreads infectiously, and a spontaneous recovery process is assumed.
Furthermore, the AVM also includes both the processes of social learning and homophilic social network rewiring that we hypothesise to be present in the empirical data.
Finally, the AVM is one of the simplest and best understood models that has these desired properties~\cite{gross2008adaptive,gross2009adaptive}.}

Specifically, the AVM considers a temporal network $\mathcal{G}(t)$ with a fixed number of $N$ nodes and $M$ links. 
Each node $v_i$ holds one of $\Gamma$ opinions or traits $o_i$ that are initially distributed at random among them. 
The $M$ links are initially distributed uniformly at random as well, thus mimicking the configuration of an Erd{\H{o}}s--R{\'e}nyi graph. 
At each discrete time step $t$, a single node $v_i$ with opinion or trait $o_i$ is randomly chosen. If its degree $k_i$, i.e.\ the number of directly connected neighbours, is non-zero, either of two processes takes place: 
\begin{enumerate}
    \item \emph{Homophilic rewiring}. With fixed probability $\varphi$ we select one of the edges that are attached to $v_i$ and move its other end to a randomly selected node $v_k$ that holds the same trait $o_k$ as $v_i$, and is not connected to $v_i$ yet. $v_i$ thereby {\em adapts} its neighbourhood structure to align more with its own trait $o_i$.
    \item \emph{Social learning}: Otherwise, with fixed probability $1-\varphi$ we pick a random neighbour $v_j$ of $v_i$ and set $v_i$'s trait equal to that of $v_j$, i.e., $v_i\leftarrow v_j$. Hence, $v_i$ {\em imitates} the trait $o_k$ of $v_k$ to become more alike to its immediate neighbourhood.
\end{enumerate}
The model reaches a steady state once only one trait per connected network component remains. In this case, no additional updates to the nodes' states or their neighbourhood structure are possible.  
The fixed probability $\varphi$ is a model parameter that allows to scale the relative frequencies of imitation and adaptation events. 
For $\varphi=0$ only imitation, and for $\varphi=1$ only adaptation takes place. 
The model displays a phase transition at intermediate values of $\varphi$ where the system's steady state qualitatively shifts from a large connected component of a single remaining trait to a \jakob{fractionalized} configuration of multiple disconnected components that each show distinct predominant traits~\cite{holme2006nonequilibrium}.  

In our specific study we set the number of nodes to %$N=850$
\niklas{$N=619$}, the number of edges to $M=5724$ and the number of traits to $\Gamma=2$ to ensure consistency with the \niklas{(filtered)} empirical data from the Copenhagen Networks Study (CNS), see below.

\subsection{Empirical temporal network data: Copenhagen Networks Study}
\label{sec:cns}

In the following, we present the Copenhagen Networks Study as our main empirical data source (Sect.~\ref{sec:data_source}) and describe the methodology used for extracting a temporal social network with time-dependent node traits from this data set (Sect.~\ref{sec:network_generation}).

\subsubsection{Description of data sources}
\label{sec:data_source}
The data analysed here originates from the Copenhagen Networks Study (CNS) \cite{stopczynski2014measuring, sapiezynski2019interaction}. 
CNS was carried out from 2012--2016 and focused on collecting temporal network and demographic data on a densely interconnected cohort of nearly 1000 individuals. 
In order to collect the temporal network information, the study handed out state-of-the-art smartphones to consenting freshman students at the Technical University of Denmark.
Specifically the study collected information on networks of physical proximity (using Bluetooth signals), phone calls, text messages, and online social networks.
In addition to the network data, the study also collected information on the participants' mobility, using the phones' GPS sensors -- and demographic and personality data, using questionnaires.
The study was approved by the Danish Data Protection agency, the appropriate legal entity in Denmark.
In terms of research, data from CNS have been used in a number of contexts e.g.\ epidemiology \cite{mones2018optimizing, stopczynski2018physical, kojaku2021effectiveness}, mobility research \cite{alessandretti2018evidence,alessandretti2020scales}, network science \cite{sekara2016fundamental, mollgaard2016measure}, studies of gender-related behaviour \cite{psylla2017role}, and education research \cite{kassarnig2017class, kassarnig2018academic}.

In addition to the data from the Copenhagen Networks Study, and in view of our aim to investigate the illustrative behaviour ``regularly going to the fitness studio'', a data set was generated with the locations of fitness studios in the vicinity of Copenhagen.
The studios were selected from the locations provided by Open Street Map \cite{OpenStreetMap} and listed with the keys 'leisure=fitness\_center' or 'sport=fitness'. A comprehensive list of all considered studios can be found in Appendix \ref{apx:locations}.

\subsubsection{Generation of empirical temporal social network}
\label{sec:network_generation}
The empirical temporal social network is generated as a physically-close-contact network between the study's participants. A network edge is created when two participants are  in close proximity to each other \jakob{once during day $t$.}
%at a time $t$. 
The network's adjacency matrix $A_{ij}(t)$ is then defined as
\begin{equation}
  A_{ij}(t) = 
  \left\{
    \begin{array}{lr}
      1\;, & \abs{s_{ij}(t)} > 80\,\text{dBm}\\
      0\;, & \text{otherwise}
    \end{array}
    \right.\;,
\end{equation}
where time $t$ is in units of days and $s_{ij}(t)$ is the maximum Bluetooth signal strength between participants $i$ and $j$ measured during day $t$, 
\jakob{while measurements where performed every five minutes}. 
The threshold $80\,\text{dBm}$ corresponds to a distance of about $2\,$m and maximises the ratio of social interactions to transient and unimportant connections \cite{sekara2014strength}.

In order to minimise noise from the beginning and end periods of data collection, 
i.e. noise due to participants joining late or dropping out early,
in this study we focus on the period from the first of February 2014 to the end of April 2014, which corresponds to the spring semester and is in the middle of the ``SensibleDTU 2013'' data collection, the second deployment of CNS.

\jakob{
% Lots of our human behaviour is repeated at regular intervals of one week. 
\jv{Much of human behaviour proceeds in weekly cycles \cite{Zuzanek1993}.
To account for}
% To capture 
this periodicity in the data, we define a time window $T(t,t')$ using a Gaussian kernel
\begin{align}
    T(t,t') &= e^{-(t-t')^2/(2t_c^2)}\;,\label{eq:time_window}\\ 
    X(t) &= \sum\limits_{t'=0}^t x(t')\cdot T(t,t')\;,\label{eq:time_window_illustration}
\end{align}
where $t_c = 7\,\text{days}$ is the characteristic time% and equation
. Eq.~\ref{eq:time_window_illustration} illustrates how $T(t,t')$ is functioning as temporal weight in a sum over an arbitrary time-dependent variable $x(t')$. 
% Variations of $t_c$ are thinkable, however, w
\jv{We suppose that} $t_c = 7\,\text{days}$ introduces the least additional assumptions as it coincides with the typical seven-day rhythm \jona{of study, work, leisure and exercise activities and behaviours (e.g. a university student would attend a particular lecture at a particular day of the week, visit the fitness study on another particular day etc.).}
The Gaussian kernel is a preferable choice to a rectangular kernel, as the latter can produce artefacts due to discontinuities. It is also a preferable choice to an exponential kernel because it decreases slowly for $t-t'<t_c$ and then tends to zero quickly. In contrast, an exponential kernel quickly falls towards zero and is therefore not suitable for a time window that represents typical horizons of human short-term activity.
}

\jakob{The raw data contains students with no or fluctuating social interaction. Reasons might be that they have left campus or spend time with people not participating in the study. 
In order to minimise their influence onto this study's results, two filters were applied to the data.
The first sorts out participants who had no or very few contacts over the whole study period by setting a lower limit for the average degree $\bar k_i \geq k_\text{min} = 4$. Variations of $k_\text{min}$ in the interval $1 \leq k_\text{min} \leq 5$ were tested, and showed no significant influence on this study's results.
The second filter compensates for the fluctuating contact behaviour of the participants. 
Some participants have a regular number of contacts on average, but occasionally this number drops to only a few or no contacts (e.g. illness could be a plausible explanation).
These absences could confound the results of the study. 
Therefore, we only consider students who had at least one contact in the last week.
For this purpose, the participants were filtered according to their average node degree in the 
% recent
\jv{past week}
%, i.e. approximately in the last week,
\begin{equation}
	\tilde k_i(t) = \frac{ \sum\limits_{t'=0}^t k_i(t') \cdot T(t,t') }
	{ \sum\limits_{t'=0}^t T(t,t') }\;.
	%\bar k_i(t) = \frac{ \sum\limits_{t'=0}^t k_i(t') \cdot e^{-(t-t')^2/(2t_k^2) }
	%{ \sum\limits_{t'=0}^t e^{-(t-t')^2/(2t_k^2) }\;,
	\label{eq:avg_n}
\end{equation}
%where
\jv{Here,} $k_i$ is the node degree and $T(t,t')$ is the time window defined in equation \ref{eq:time_window}.
We therefore interpret
$\tilde k_i(t)$ as the average number of daily contact events in past week, and we
consider only students in our analysis that had in the order of one contact in the last week,
i.e. we set the lower bound to $\tilde k_i(t) \geq \tilde k_{\text{min}} = 1/7$.
Variations of $\tilde k_\text{min}$ in the interval $1/7 \leq k_\text{min} \leq 1$ were tested, and showed no significant influence on this study's results.}

In order to investigate possible spreading dynamics of the illustrative behaviour ``regularly going to the fitness studio'', 
we match stop-locations with the locations of fitness studios (Appendix \ref{apx:locations}). Here, stop-locations are coordinates generated from the GPS data, where the participants spent at least 15 minutes \cite{Cuttone2014}. The accuracy chosen for matching is $10\,\text{m}$, which corresponds to the precision of GPS \cite{gps2008techreport}.
Hence, we record for each node $i$ at the time $t$ the behaviour
\begin{equation}
    b_{i}(t) = 
			\left\{ 
				\begin{array}{lr} 
				    1\;, & \text{if node } i \text{ visited a studio at day } t \\ 
				    0\;, & \text{otherwise}
				\end{array} 
			\right.\;.
\end{equation}
To distinguish between students who go to the studio occasionally and students who go regularly, we introduce the %smoothed behaviour %:
\jakob{past-week behaviour}
\begin{equation}
    %\bar{b}_i(t) = \sum\limits_{t'=0}^t b_i(t') \cdot e^{-(t-t')^2/(2t_b^2)}\;,
    \jakob{\bar{b}_i(t) = \sum\limits_{t'=0}^t b_i(t') \cdot T(t,t')\;,}
\end{equation}
\jakob{with $T(t,t')$ the one-week time window defined in equation \ref{eq:time_window}. We interpret $\bar b_i(t)$ as typical behaviour during the last week.}
%with the characteristic time $t_b = 7\,\text{days}$. The one-sided Gaussian kernel $e^{-(t-t')^2/(2t_b^2)}$ is chosen to favour current behaviour occurring close to time $t$, where the kernel reaches values close to one. Conversely, it suppresses past behaviour. 
%Thus, $\bar b_i(t)$ can be interpreted as the typical behaviour in the last seven days.

\begin{figure}[h]
    \centering
    \includegraphics[width=\scale\textwidth]{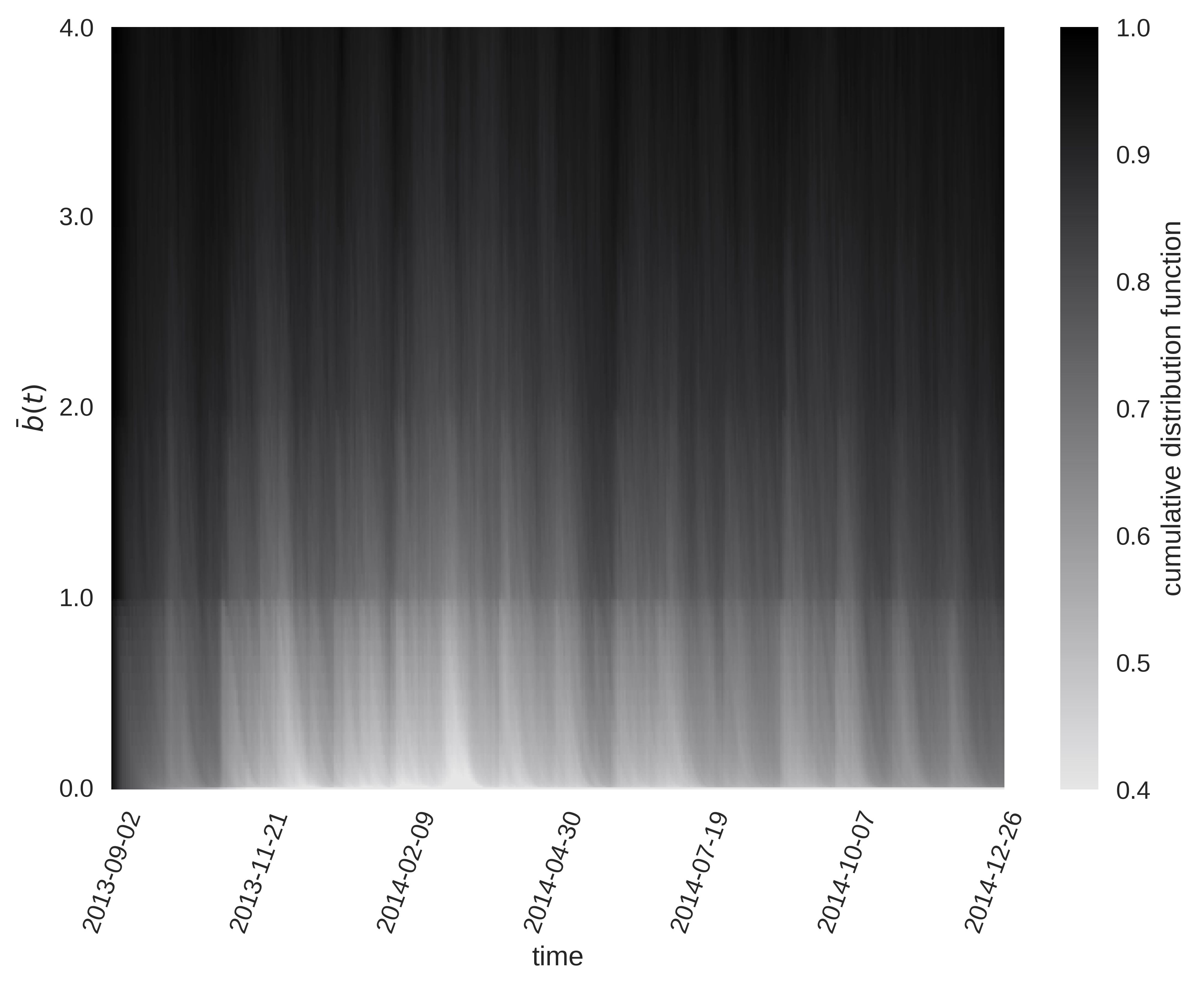}
    \caption{Cumulative distribution of the 
    %smoothed behavioural 
    \jakob{past-week behavioural}
    function $\bar b(t)$ plotted as a heat map over the period of the entire ``SensibleDTU 2013'' data collection. Our study analyses the three month subperiod from February to April 2014. A clear boundary is visible at $\bar b(t) \approx 1$ for all $t$, with values of $\bar{b}(t)>1$ being much less frequent than $\bar{b}(t)<1$. Therefore, $\bar b(t) = 1$ is a reasonable choice to separate the participants into two groups. Members of the group with $\bar{b}(t)\geq 1$, 
    \jakob{who visited the fitness studio at least once in %approximately 
    the past week}
    %who visit the fitness studio at frequent intervals,
    are referred to as active nodes, while individuals with $\bar{b}(t)<1$ are referred to as passive nodes.}
    \label{fig:cdf}
\end{figure}

Finally, for each point in time $t$ we split the participants into two groups: 
(i) students going occasionally or not at all 
% \jakob{$\bar b_i(t) < \gamma$ in the past week}
to the fitness studio, and 
(ii) students going more often
% \jakob{$\bar b_i(t) \geq \gamma$ in the past week}
to the studio.
\jv{A typical behaviour of regularly going into the fitness studio would be to go once a week. This suggests to select $\bar b_i(t)  = 1$ as 
% an intuitive threshold choice.
a threshold criterion,
and to explore the following}
% and a
time-dependent trait $o_i(t)$ for each node in the network, %:
\begin{equation}
    o_i(t) =  
	    \left\{ 
	        \begin{array}{lr} 
			    1\;, & \bar{b}_i(t) \geq 1 \\ % \gamma\\ 
			    0\;, & \text{otherwise}
			\end{array} 
		\right.\;.
\end{equation}
\jakob{
% A typical behaviour of regularly going into the fitness studio would be to go once a week. Hence, $\gamma = 1$ is an intuitive threshold choice. 
% Moreover
\jv{Indeed, there is}  % we can observe 
a clear boundary in the cumulative distribution of $\bar b(t)$ plotted in Fig.~\ref{fig:cdf} for $\bar b(t) \approx 1$ and for all $t$.
The boundary indicates that $\bar{b}(t)>1$ is occurring less frequently than $\bar{b}(t)<1$. This supports the choice to separate participants with the threshold $\bar b(t) = 1$.
In the following, the students going to gyms at least once in the last week~($o_i(t) = 1$) are referred to as ``active'' nodes, while the others~($o_i(t) = 0$) are referred to as ``passive'' nodes.
}
%As threshold, $\gamma = 1$ is chosen, motivated by a clear edge in the cumulative distribution of $\bar b(t)$ plotted in Fig.~\ref{fig:cdf}. 
%The edge is visible at $\bar b(t) \approx 1$ for all $t$, with values of $\bar{b}(t)>1$ occurring less frequently than $\bar{b}(t)<1$ . 
%This suggests that it is reasonable to separate participants between those who go to gyms regularly $\gamma \geq 1$, and those who go only occasionally $\gamma \leq 1$. The former will be referred to as ``active'' nodes, and the latter as ``passive'' nodes.

\jakob{The procedure presented here generates a social network consisting of $619$ nodes with an average degree of $\bar{k}_i = 19$. The nodes change their trait $o_i(t)$ on average $5.94$ times} \marc{over the course of the considered three months-period.}

\section{Methods}
\label{sec:methods}
In this section, we describe the methodologies used to estimate empirical dose response functions from temporal network data (Sect.~\ref{sec:dose_response_fct}) and for generating surrogate data sets to test hypothesis on the processes and structures underlying specific features of the empirical dose response functions (Sect.~\ref{sec:surrogate_data}).

\subsection{Estimating dose-response functions from temporal network data}
\label{sec:dose_response_fct}
Dose response functions (DRFs) represent the functional dependence between the probability of changing a trait $p_{o\rightarrow o'}$ and the exposure $K$, which is defined as the joint influence of all contacts with a given trait, or more formally as the superposition of all received doses from neighbouring nodes. 
\jakob{
We assume that the influence of each node is equal and that the recent influence from the last week has a greater impact on the decision-making process than the influence from the distant past, i.e. it contributes more to the exposure $K$. 
}
To measure the exposure to which a single node $i$ is subjected, we put

\begin{equation}
	%K_i(o,t) = \sum\limits_{t'=0}^t \mathcal{N}_i(o,t') \cdot e^{-(t-t')^2/(2t_K^2)}\;,
    \jakob{K_i(o,t) = \sum\limits_{t'=0}^t \mathcal{N}_i(o,t') \cdot T(t,t')\;,}
	\label{eq:absolute_exposure}
\end{equation}
where $\mathcal{N}_i(o,t')$ is the number of neighbouring nodes with trait $o$ at time $t'$ 
\jakob{and $T(t,t')$ is the weight of the encounter as defined in Eq. \ref{eq:time_window}, which 
% devalues
\jv{down-weights the}
influences from 
\jv{encounters}
% distant past, i.e. 
from further back than one week.
}
%Hence, we assume that each node's influence is equal. 
%The one-sided Gaussian kernel $e^{-(t-t')^2/(2t_K^2)}$ together with the characteristic exposure time of $t_K = 7\,$days acts as a smoothing. Contacts in the near past $t-t' \lesssim t_K$ dominate the sum due to the weighting by the kernel. Conversely, contacts in the distant past $t-t' \gtrsim t_K$ are devalued. 
%We thus interpret the kernel as representing the memory capacity of node $i$ for the period of approximately $t_K = 7\,$days.

From the time series of each node's traits $o_i(t)$, the received exposures $K_i(o,t)$ can be computed,
allowing us to estimate the DRFs as relative frequencies as
\begin{equation}
   \p(K) \approx  \frac{C(K)}{N(K)}\;.%,
   \label{eq:pswap}
\end{equation}
Here $C(K)$ is the number of nodes that have changed their trait between $t-1$ and $t$ and having experienced a certain level of exposure~$K$. Furthermore, $N(K)$ is the total number of nodes that have experienced exposure level $K$. $C(K)$ and $N(K)$ are the result of an aggregation over all time steps and are thus time-independent.

$p(K)$ is an estimator of the actual probability of changing trait when experiencing an exposure level of $K$. 
If the reactions (changing trait or not) to subsequent exposures are assumed to be independent, this estimator is simply the empirical success rate of an $N(K)$ times repeated Bernoulli experiment, and its standard error can thus be estimated by
\begin{equation}
    %\niklas{\sigma_p = %\sqrt{\frac{C(K)(N(K)-C(K))}{N(K)}}\;.}
    \jobst{
    \sigma_p =
    \sqrt{\frac{p(K)(1-p(K))}{N(K)}}
    =
    \sqrt{\frac{C(K) \bigl( N(K)-C(K) \bigr)}{N(K)^3}}\;.
    }
\end{equation}
In the present study we \jv{adopt} %ed
%$\sigma_p^c=\sqrt{C(K)(N(K)+C(K))}/N(K)$
\niklas{\begin{equation}
\sigma_p^c=\sqrt{C(K) \, \bigl(N(K)+C(K)\bigr) /N(K)^3}\label{eq:error-bars}\end{equation}}
as a conservative upper bound to this error.
\niklas{Where multiple data sets are used for one result, as is the case when multiple simulation runs or surrogate model realisations are computed using the same parameters, the data are considered as one ensemble for further analysis. 
The error estimation in Eq.~\ref{eq:error-bars} is thus performed on these pooled data sets where applicable.}

\subsection{Generating surrogate data sets for hypothesis testing}
\label{sec:surrogate_data}
%

%\begin{figure}[h]
%    \centering
%    \includegraphics[width=\%scale\textwidth]{figures/sur%rogate-work-flow.png}
%    \caption{Mock-up: %Overview of the surrogate %data model methodology used %for detecting spreading %processes in temporal %network data.}
%    \label{fig:surrogate-met%hod}
%\end{figure}

To probe the empirical data from the Copenhagen Networks Study for contagion effects relating to the studied behaviour, we use the method of surrogate data sets. 
The surrogate data approach is a statistical method for identifying non-linearity, such as contagion effects, in time series. 
This is achieved by performing hypothesis tests on data sets that are generated from the empirical data by using Monte Carlo methods \cite{theiler_testing_1992, schreiber_surrogate_2000, gauvin_randomized_2020, holme_temporal_2012}. 
Surrogate data sets have been used in the past to study a wide range of time series \cite{venema_statistical_2006, scheinkman_nonlinear_1989, pritchard_dimensional_1995} and network data \cite{wiedermann_spatial_2016, maslov_detection_2004, maslov_specificity_2002}.
The method is described in the following paragraph, followed by the description of the surrogate data studies 
% presented in this
\jv{examined in the present}
contribution.

First, a class of processes that may potentially be sufficient in explaining the empirical data, is specified as a composite null hypothesis $\mathcal{H}_0$.
To test this hypothesis, a new, ``surrogate’’ data set is derived from the empirical data in a way that is consistent with $\mathcal{H}_0$. 
Any structures that the null-hypothesis excludes are destroyed in this process, while other features of the original data are retained.

One algorithm which can be used to produce such surrogate data sets is the creation of random permutations of the original data, for example by permuting the nodes' time series or network connections.
The product resembles the empirical data, but lacks the features excluded by the null hypothesis, such as contagion processes. 
This method, known as Constrained Realisations \cite{theiler_constrained-realization_1996}, represents a parameter-free way of producing surrogate data sets without the use of a specific model.
A discriminating statistic is then computed on the original data and surrogate data sets alike. If there is a significant difference between the value or distribution computed for the original data, and the ensemble of values or distributions computed for the surrogate data sets, the null hypothesis is rejected. 
Put simply, the empirical data are permuted in a way that is consistent with a composite null hypothesis, and if this substantially changes a statistical measure of interest, the null hypothesis can be rejected.
Through the careful choice of iteratively more complex null hypotheses, preserving different sets of data properties, the nature of the true underlying non-linear process can be investigated. 

Six surrogate data sets are produced for this analysis. The first four investigate the influence of different assumptions about the node dynamics on the dose response functions, by permuting the node traits $o_i(t)$ and keeping the network component $A_{ij}(t)$ unchanged. 
The last two surrogate models address the effect of the network component, by permuting the network edges $A_{ij}(t)$ and keeping the node dynamics $o_i(t)$ unchanged.
\niklas{An overview of the investigated null hypotheses is displayed in Fig.~\ref{fig:surrogate_overview}B. 
In this figure, arrows from a surrogate test at a higher to one at lower location indicate a higher degree of randomisation in the former than in the latter.
This illustrates the hierarchical nature of surrogate randomisation models.
To describe the surrogate data sets $P$ associated with the null hypotheses $\mathcal{H}_0$, the canonical naming convention from \cite{gauvin_randomized_2020} is used.
This convention is based on defining surrogate data sets by the quantities they conserve with respect to the original data.}
In the following, the estimated DRF of the empirical data is referred to as the empirical DRF $\p$, while the one estimated for surrogate data may be referred to as the surrogate DRF $\ps$.
\niklas{To reduce statistical uncertainties, ten surrogate data realisations are performed for each null hypothesis.
% These data sets are pooled before the computation of
\jv{They are considered as one ensemble to compute}
the dose response functions and their error bars.}
The following surrogate data test were conducted:

\begin{enumerate}
\item \textit{$\mathcal{H}_0^1$: $\niklas{P(A_{ij}(t), O)}$. 
The empirical DRF can be reproduced with a class of models that is based only on the global mean activity level $O=\overline{\langle o_i(t)\rangle_i}$.} 
Here, the overline and brackets represent the time and ensemble average, respectively. 
This null hypothesis represents the most basic assumption, corresponding to an underlying process that is completely random. 
For this surrogate data set, all traits $o_i(t)$ are permuted randomly. 
Only the average activity level across the entire ensemble and observation period is conserved. 
\item \textit{$\mathcal{H}_0^2$: $\niklas{P(A_{ij}(t), O_i)}$.
The empirical DRF can be reproduced with a class of models that is based only on each node’s individual activity level $O_i=\overline{o_i(t)}$.} 
This null hypothesis leaves room for an activity factor unique to each individual node, while still assuming otherwise random node dynamics. 
For the corresponding surrogate data set, the activity levels are permuted in time, separately for each node.
\item \textit{$\mathcal{H}_0^3$: $\niklas{P(A_{ij}(t),\{\tau_{i;0,1}\})}$. 
The empirical DRF can be reproduced with a class of models
\niklas{that is based only on the distribution of time intervals for which the node stays in either activity state $\tau_{i;0,1}$, which implicitly conserves $O_i$ and the number of activity level switches as well.} }
%that is based only on each node’s individual activity level $O_i$, \jobst{its number of activity state switches, and the distribution of time intervals for which the node stays in either activity state} \niklas{ $\{\tau_{i;0,1}\}$}}. 
This null hypothesis builds on the previous one by also conserving each node’s \jobst{overall} persistence, defined as the inverse of a node's number of switches between behaviours\jobst{, and the corresponding distribution of time intervals}. 
This is realised by %separately 
permuting the length of intervals with a constant activity level, separately for periods of active and passive behaviour, for each node. \jobst{E.g., the sequence (active for 2 steps, inactive for 5 steps, active for 3 steps, inactive for one step) may be turned into (active for 3 steps, inactive for one step, active for 2 steps, inactive for 5 steps).}
\niklas{The number of activity level switches is a constraint on the randomisation space for this surrogate model.
However, the average number of activity level switches allows for sufficient randomisation in our data (see Appendix \ref{apx:h03}).}
\item \textit{$\mathcal{H}_0^4$: $\niklas{P(A_{ij}(t), O(t))}$.
The empirical DRF can be reproduced with a class of models that is based only on the mean time-dependent activity level $O(t)=\langle o_i(t)\rangle_i$ of the ensemble.} 
This null hypothesis assumes a non-stationary temporal dynamics of the ensemble’s behaviour, while excluding any non-random individual node characteristics. 
The surrogate data set is produced by permuting the activity states of all nodes, separately for each time step.
\item \textit{$\mathcal{H}_0^5$: $\niklas{P(A, O_i(t))}$.
The empirical DRF can be reproduced with a class of models that is based only on individual activity dynamics and the average network edge density $A=\overline{\langle A_{ij}(t)\rangle_{i,j}}$.} 
In this case, the null hypothesis contains the assumption that the observed DRF is independent of the specific topology of the connection network, and arise solely based on the individual nodes’ behaviour. 
The corresponding surrogate data set is produced by randomly permuting all edges across nodes and time.
\item \textit{$\mathcal{H}_0^6$: $\niklas{P(k_i(t), O_i(t))}$.
The empirical DRF can be reproduced with a class of models that is based only on the individual node dynamics, and each node's time-dependent network degree $k_i(t)=\sum_{j=0}^N A_{ij}(t)$.} This null hypothesis builds on the previous one by randomising the neighbourhood of the nodes, but preserving each nodes connectivity in the network. This can serve as a check for homophilic effects in the network dynamics. To produce the surrogate data set, we use the random link switching algorithm \cite{zamora-lopez_reciprocity_2008, artzy-randrup_generating_2005}. Pairs of connections $(i,j)$ and $(k,l)$ are drawn randomly, and are transformed into the connections $(i,k)$ and $(j,l)$. This procedure ensures that each node's degree remains unchanged.
\end{enumerate}

We choose the dose response function, introduced in Sect.~\ref{sec:dose_response_fct}, as the discriminating statistic used to compare empirical and surrogate data sets. 
The comparisons of surrogate DRFs $\pas$ and empirical $\pa$ DRFs are presented in Sect.~\ref{sec:results_empirical}.
\niklas{To test our methodology, we also create the hierarchy of surrogate models for the synthetic AVM data with realistic parameter choices \jv{(see Appendix~(\ref{apx:AVMsurrogates})).}}
\jakob{In order to quantify the difference between $\pas$ and $\pa$,} 
\jobst{we use a test statistic $\zeta$ that combines the $k$ many individual $z$-scores 
\jakob{(denoted as $z_i, i=1,...,k$)} 
of the DRFs into a single score similar to Stouffer's z-score method \cite{Stouffer1949, WHITLOCK2005}, but using the sum of squared $z$-scores instead of their simple sum so that negative and positive deviations cannot cancel out. Since under the null hypothesis, that sum has a $\chi^2$-distribution with $k$ degrees of freedom, which depends in a nontrivial way on $k$, we additionally normalize the sum of squares by dividing it by the 95th percentile of that distribution, so that a value of $\zeta\ge 1$ indicates a significant deviation from the null hypothesis: 
\begin{equation}
    \zeta = \sum_{i=1}^k z_i^2 / Q_{0.95}(\chi^2_k).
    \label{eq:zeta_score}
\end{equation}
} 
\section{Results}
\label{sec:results}
Here, we report on the results obtained by applying our proposed dose response function methodology. As a first step, we analyse synthetic data generated by the adaptive voter model as a proof of concept (Sect.~\ref{sec:results_avm}). Building on these insights, we then investigate the empirical temporal network data obtained from the Copenhagen Network Study (Sect.~\ref{sec:results_empirical}).
Our findings are summarised in Sect.~\ref{sec:summary}.

\subsection{Synthetic data}
\label{sec:results_avm}
\begin{figure}[h]
	\centering
	\includegraphics[width=0.495\linewidth]{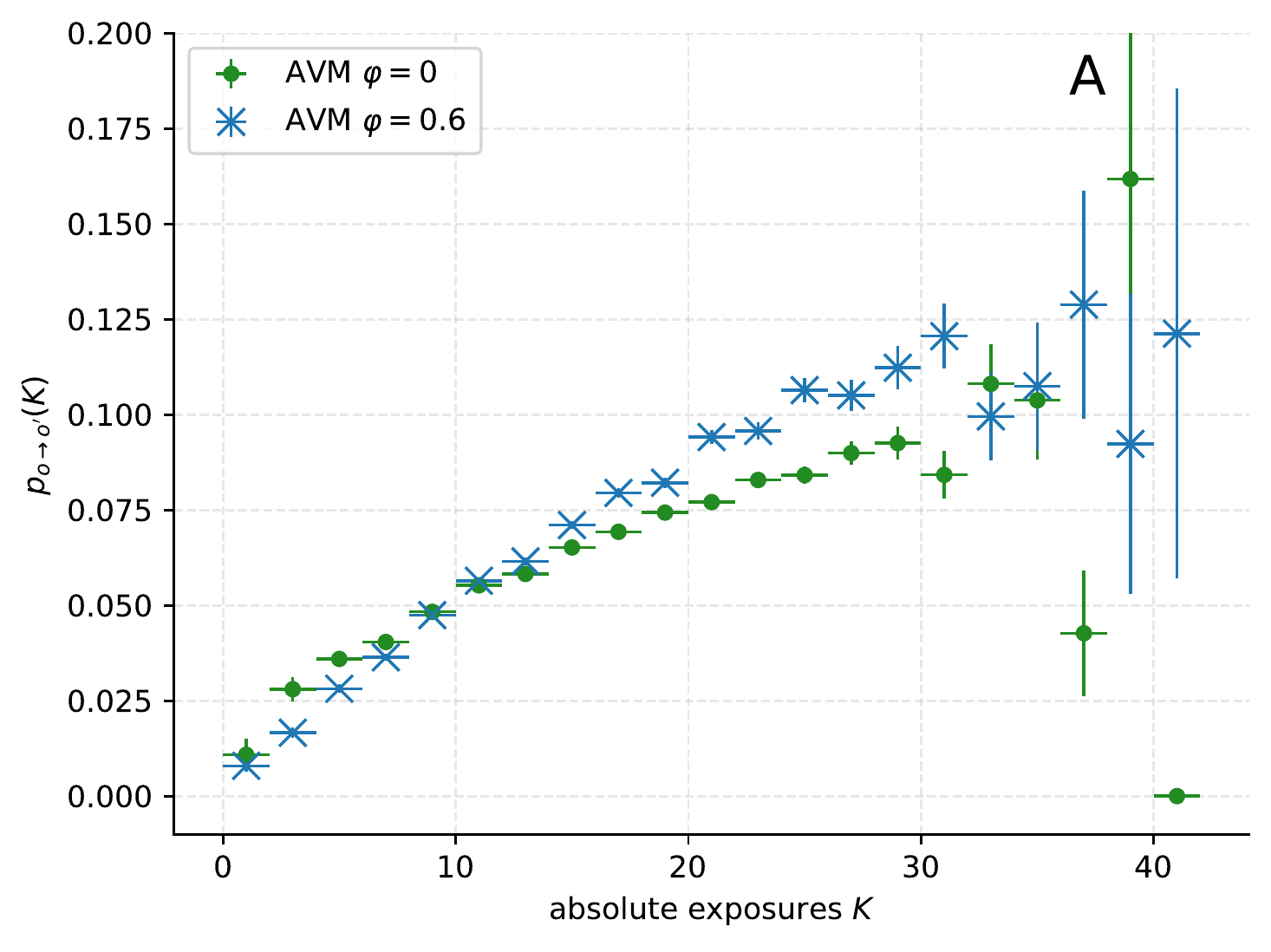}
	\includegraphics[width=0.495\linewidth]{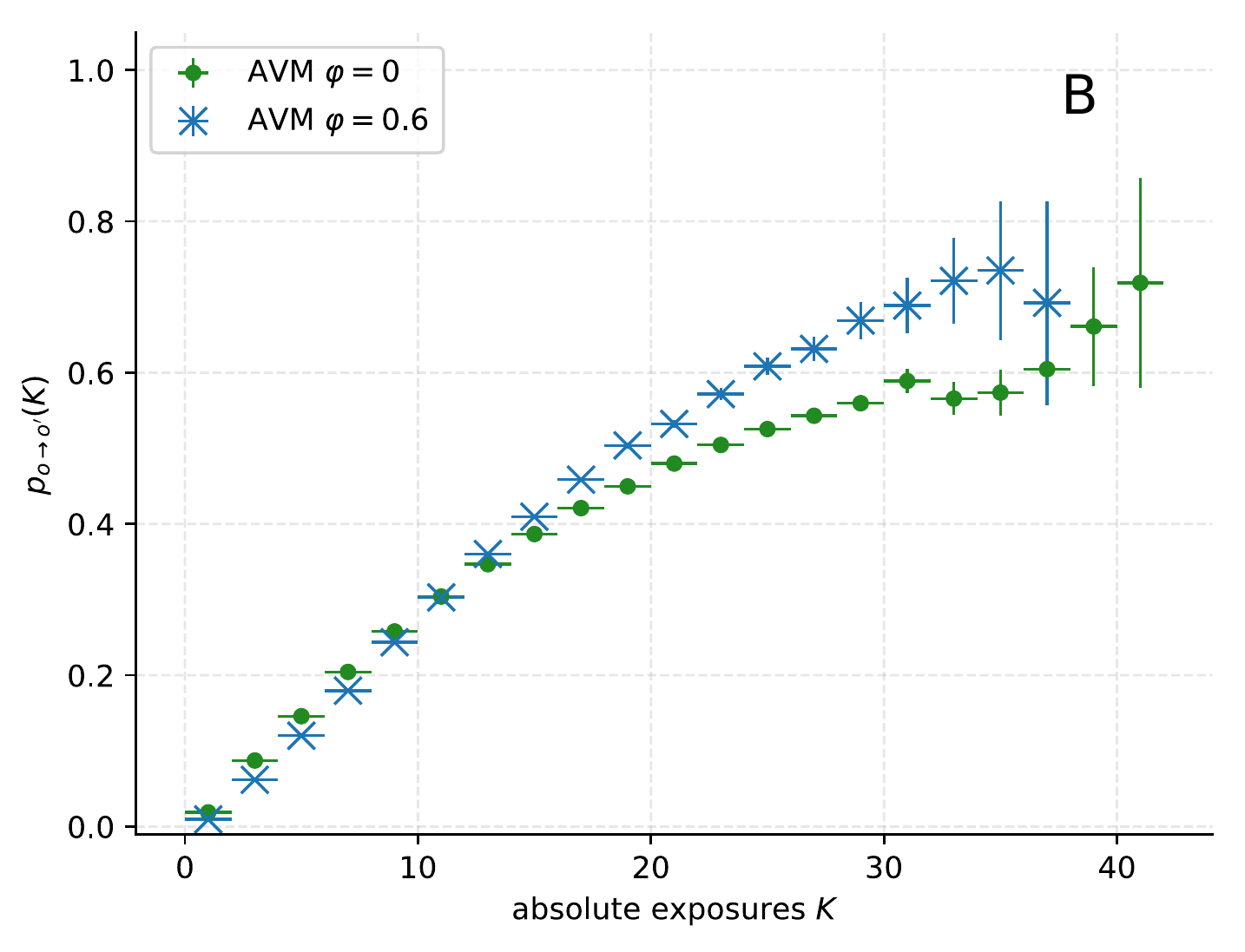}
	\caption{
	\jakob{
	Average estimated dose response functions (DRFs) for synthetic temporal network data generated by ten runs of the adaptive voter model for rewiring probability $\varphi = 0$ and $\varphi = 0.6$. 
	\niklas{Error bars are computed as described in Sect.~\ref{sec:dose_response_fct}.}
	%Error bars are based on a conservative estimate of the standard error of the switching probabilities $p(K)$ (Eq.~\ref{eq:error-bars}), based on the combined data set of all model simulation runs. 
	In (A) the number of nodes $N=619$, the average degree $\bar k_i = 19$ and the number of simulated time steps $\tau = 90$ were chosen analogously to the empirical temporal network from the Copenhagen Networks Study. % In particular, t
	\niklas{T}he number of model updates per time step was adjusted to align with the average number of behaviour changes per time step with the CNS data\niklas{.}% as well. 
	The error for the data point at $K\approx 41$ could not be estimated due to a lack of measurements; a large error is plausible.
	In (B) the simulations are repeated for a larger network with $N=851$ nodes, an average degree of $\bar k_i = 13.5$ and significantly more model updates per time step.
	%Furthermore, the whole simulation from initialisation until consensus 
	\niklas{Here, the system was simulated until consensus} (i.e. all nodes having the same trait for $\varphi = 0$, while for $\varphi = 0.6$ the model converges to two distinct groups with consensus each) % with $\tau = 190$ steps was analysed.
	\niklas{was reached at $\tau = 190$ steps.}
	Both (A) and (B) show a monotonic increasing relationship between $\pa$ and $K$, while in (B) this trend is clearer due to the larger number of data points.
	The DRFs for $\varphi = 0$ differ significantly from those for $\varphi = 0.6$, owing to the more progressed network adaptation in the latter case.
	This difference shows that their form is not only influenced by contagion (imitation or social learning) effects, but also by homophily (network adaptation) dynamics.
	}
	}

	\label{fig:DRF_example}
\end{figure}

As a first application of our methodology, we analyse synthetic temporal network data generated by the adaptive voter model (Sect.~\ref{sec:avm}). 
Fig.~\ref{fig:DRF_example} shows the estimated DRFs for the AVM with $\varphi = 0$ (green dots), which includes only imitation dynamics, and with $\varphi = 0.6$ (blue crosses), involving both imitation and homophily dynamics. 
\niklas{Two cases are simulated: In Fig.~\ref{fig:DRF_example}A, model parameters are chosen to align the average frequency of behaviour switches across the system, and the number of time steps, with the data from the CNS study. 
To display the effects of more progressed network adaptation, Fig.~\ref{fig:DRF_example}B displays the DRF of a similar simulation, where the model updates per time step, and the total number of simulated time steps, are significantly increased.}
\jv{Each plot contains}
% The plots contain the 
data from ten independent model runs. % each. 
The probabilities for the change of trait $\p$ are generated for equally sized bins with a width of $K=2$. 
Only bins with at least 30 data points were considered.
\jv{For increasing}
% Nevertheless, for high 
$K$, the DRF $\p$ is subject to increasing uncertainties, since exposures $K>30$ are very rare in the network.

As suggested by the imitation rule in the model, we observe that $\p$ depends monotonically, but non-linearly, on $K$.
Moreover, the plots for $\varphi = 0.6$ 
% \niklas{in Fig.~\ref{fig:DRF_example}B} 
clearly show the impact on $\p(K)$ of the additional homophily compared to the plot of $\varphi = 0$. 
For $K \gtrsim 15$ the DRF of this data is significantly larger then for those with $\varphi = 0$.
% \niklas{The difference between the two DRFs is much less prominent in Fig.~\ref{fig:DRF_example}A, where the effects of network adaptation are only beginning to take shape.
% This also manifests in the smaller range of values for $K$: Since the node neighborhoods have not been adapted as much, they remain more heterogeneous, and hence very high exposures are much rarer. }
\jv{For $K\gtrsim 30$ the difference between the DRFs is obscured by the increasing errors in case A, but it is still clearly showing for the longer simulations in panel B.}

From this first proof of concept application, we can conclude that contagion dynamics such as the imitation rule in the model \cite{Dodds2004universalBehavior,Dodds2005generalizedModel} leads to positive correlation of $\p$ and $K$. 
However, from the estimated DRF for $\varphi =0.6$, we learn that homophily is reflected in the DRFs as well. 
To distinguish between the different dynamics, we use a surrogate analysis in the following investigation of the empirical temporal network data (Sect.~\ref{sec:surrogate_data}). 

\niklas{To validate our data analysis methodology, we \jv{computed} the complete hierarchy of surrogate models (described in Sec.~\ref{sec:surrogate_data}) on the synthetic AVM data set with CNS-aligned parameter choices.
The details of this study are given in Appendix \ref{apx:AVMsurrogates}, while the results are summarised in Fig.~\ref{fig:surrogate_overview}A.
In line with our expectations, we 
% found 
\jv{find}
evidence for contagion effects in both the $\varphi=0.0$ and $\varphi=0.6$ cases.
Significant homophilic effects 
% were
\jv{are} 
only found where the network adaptation process of the AVM was active ($\varphi=0.6$), also confirming our expectations.
This demonstrates the sensitivity and appropriateness of our methodology for detecting contagion and homophily in the studied empirical data set.}
\jv{A detailed exposition of the approach is now given for the empirical data on the Copenhagen network study. 
Subsequently, the results for both the synthetic and the empirical data are discussed in Sect.~\ref{sec:summary}.}

\subsection{Empirical data}
\label{sec:results_empirical}
\begin{figure}[ht]
	\centering
	\includegraphics[width=\Lscale\linewidth]{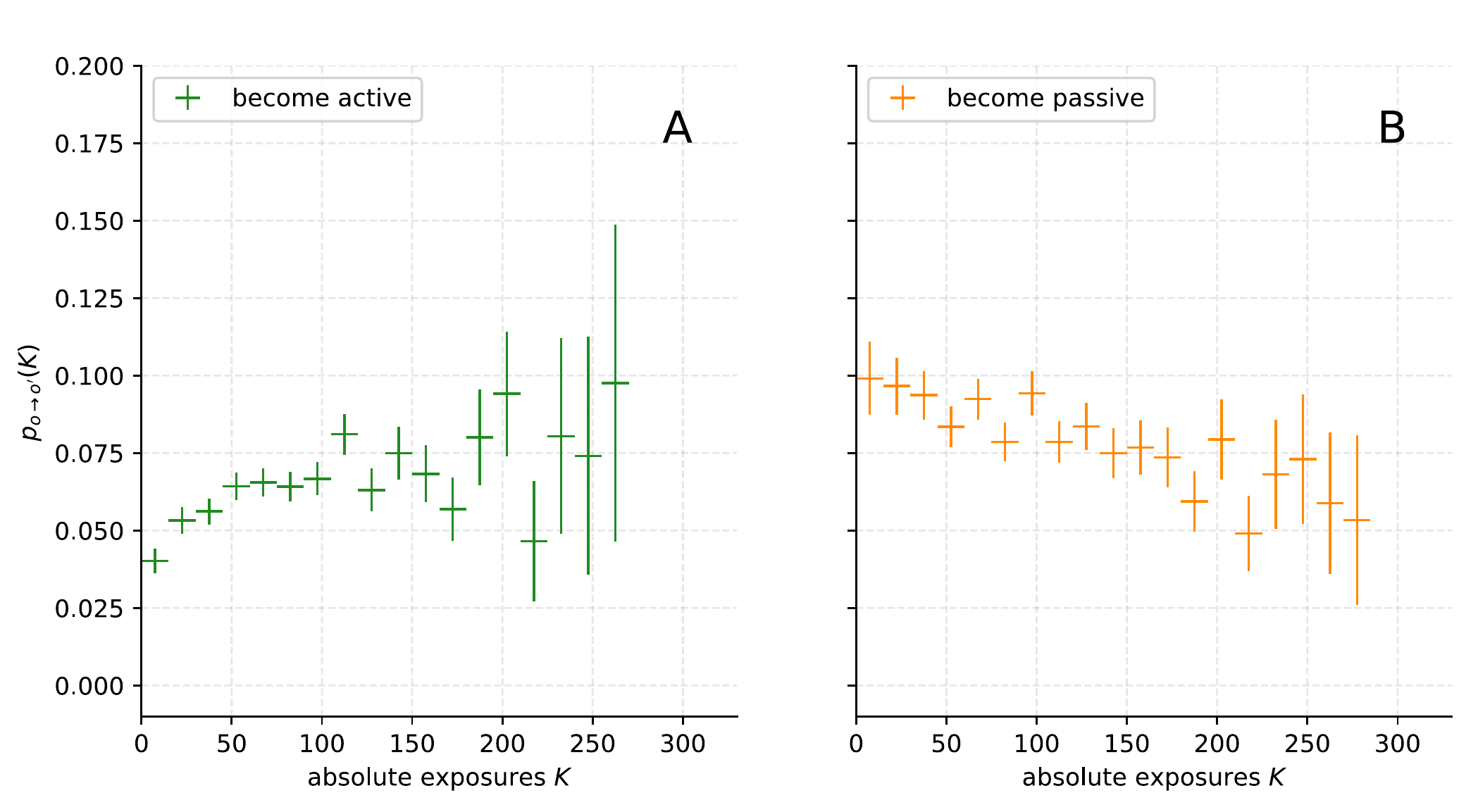}
	\caption{Empirical dose response functions computed from the Copenhagen Networks Study temporal network data, representing the probability to become active (A) or passive (B), as a function of the absolute exposure to these respective activity levels. 
	%\jona{Error bars for the empirical DRFs are based on a conservative estimate of the standard error of the switching probabilities $p(K)$ (Eq.~\ref{eq:error-bars}).} 
    \niklas{Error bars are computed as described in Sect.~\ref{sec:dose_response_fct}.}
	For the probability to become active $\pa$, a clear upward trend is noticeable \jakob{($\rho = 0.61$; $p=0.007$)}, which might be caused by contagion.
	%, although the sparse data at $K>170$ make it difficult to discern this trend there.
	For the probability to become passive $\pp$, \jakob{a monotonic decrease can be identified ($\rho = -0.89$; $p=3.5\cdot 10^{-7}$).}
	%no clear dependence on $K$ can be identified due to large uncertainties. 
	%Note that the two estimated DRFs are very different from those derived from the adaptive voter model shown in Fig.~\ref{fig:DRF_example}.
	}
	\label{fig:DRF_abs_expo}
\end{figure}

In the following, we apply our methodology to empirical temporal network data from the Copenhagen Networks Study (Sect.~\ref{sec:cns}) to investigate possible spreading dynamics of the illustrative behaviour ``regularly going to the fitness studio''.
The DRF $\p(K)$ is estimated for equal-sized bins with a width of $K=5$. Only bins with at least 30 data points were considered. The resulting DRFs are shown in Fig.~\ref{fig:DRF_abs_expo}.

We observe that the probabilities for becoming active $\pa$ (Fig.~\ref{fig:DRF_abs_expo}A) and for becoming passive $\pp$ (Fig.~\ref{fig:DRF_abs_expo}B) do not behave in a symmetric way. 
Since the initiation and the maintenance of an activity represent two rather distinct phases \cite{marcus_transtheoretical_1994}, this is not necessarily surprising.
%For the latter, $\pp$, a slight negative dependence on $K$ may be indicated, however this is obscured by the large error bars.
\jakob{
To test whether we observe significant monotonic relationships of $\pa(K)$ and $\pp(K)$ with $K$, we calculate Spearman's rank correlation coefficient $\rho$ \cite{Spearman1904}. For a perfect monotonic increase (decrease), the coefficient is equal to $\rho = 1$ ($\rho = -1$), while $\rho = 0$ indicates the absence of a monotonic relationship.
}
\jakob{For $\pp$ a slight but significant monotonic decrease can be identified with $\rho = -0.89$ and a $p$-value of $p=3.5\cdot 10^{-7}$.}
%Stopping to regularly go to the fitness centre could possibly be largely independent of contagion events and dominated by external influences (e.g.\ an injury). Therefore, in the following we focus our analysis on the probability of becoming active $\pa$.
\jakob{Going to the gym more often than contacts (large $K$) could potentially be an incentive to maintain active behaviour and lead to the observed monotonic decrease. However, we address in this study the switching between active and passive behaviour as a consequence of social contagion and therefore focus on the probability of becoming active $\pa$ in the following analysis.}

The probability $\pa$ is subject to large errors for $K>100$. The low occurrence of large $K$ seems to be the main reason. 
%However, we find a notable positive correlation of $\pa$ with $K$ for $K<100$, which could indicate contagion or homophilic dynamics.
\jakob{However, we find a significant monotonic increase of $\pa$, with Spearman's rank correlation coefficient $\rho = 0.61$ and $p$-value $p=0.007$. This correlation could indicate contagion or homophilic dynamics.}
To pursue this indicator further, we examine the DRF using the surrogate data set method (Sect.~\ref{sec:surrogate_data}). First, we investigate the possible influence of contagion dynamics (Sect.~\ref{sec:investigation_contagion}), then for group dynamics or external influences (Sect.~\ref{sec:investigation_group}) and finally for homophily dynamics (Sect.~\ref{sec:investigation_homophily}).

\subsubsection{Investigation for Contagion Dynamics}
\label{sec:investigation_contagion}

For investigating the possible influence of contagion dynamics on the DRF we 
employ the surrogate data tests 
$\mathcal{H}_0^1$,
$\mathcal{H}_0^2$, and
$\mathcal{H}_0^3$
introduced in Sect.~\ref{sec:surrogate_data},
i.e., consider surrogate models in which explicitly no contagion takes place
and we explore if they nevertheless reproduce the empirically observed DRF.
To do so, we permute the traits of the nodes $o_i(t)$ and leave the network component $A_{ij}(t)$ unchanged. These permutations destroy possible temporal correlations of exposure $K$ with changes in traits and, thus, any trace of contagion dynamics.
In three steps, we analyse the impact of different assumptions about the node dynamics on the dose-response functions and show step by step which assumptions are necessary to explain the observed DRF. 

\paragraph{First Data Test. Hypothesis $\mathcal{H}_0^1$: $\niklas{P(A_{ij}(t), O)}$.} 
\textit{The empirical DRF can be reproduced with a class of models that is based only on the global mean activity level $O=\overline{\langle o_i(t)\rangle_i}$.} 

\begin{figure}[h]
	\centering
	\includegraphics[width=\Lscale\linewidth]{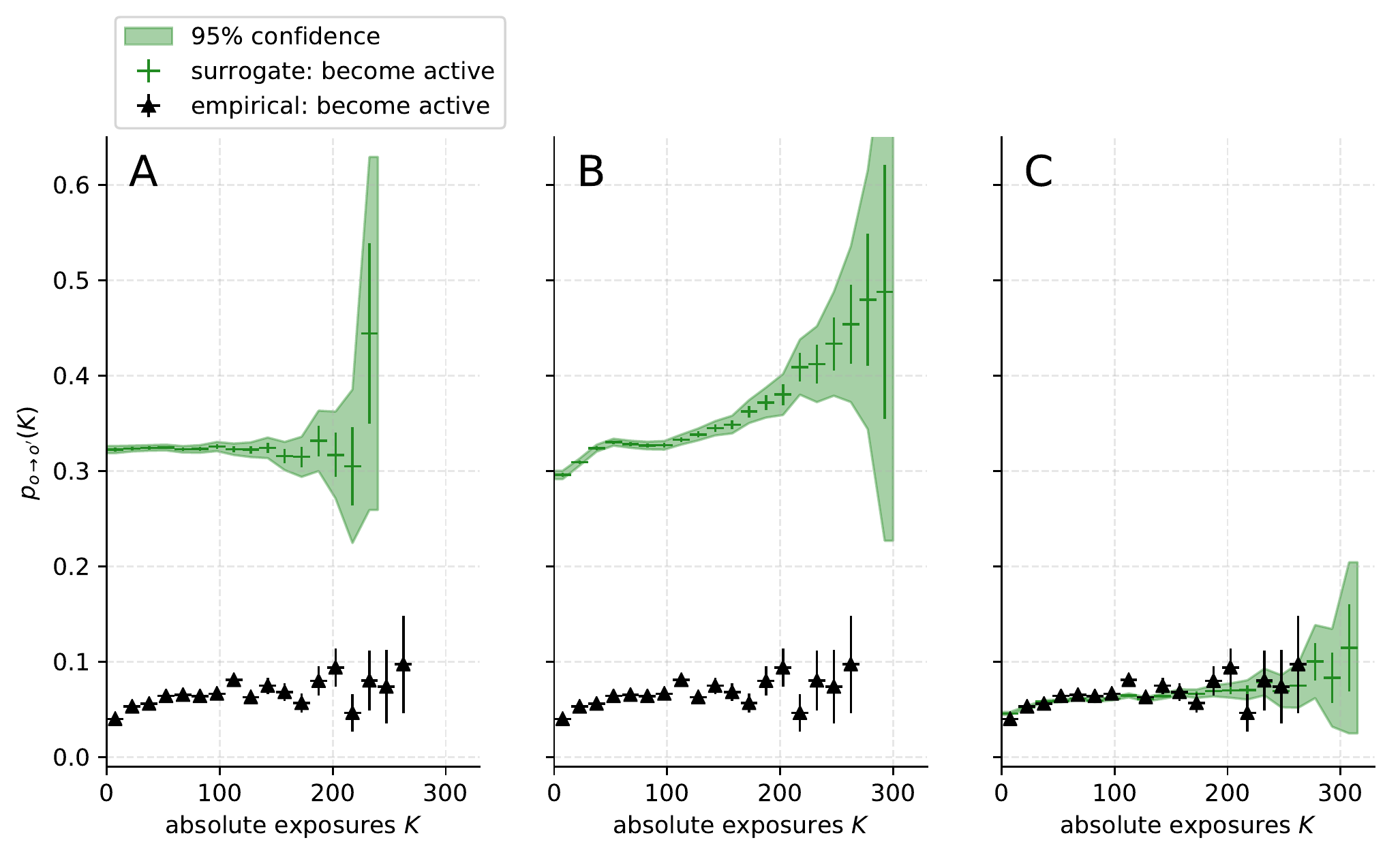}
	\caption{Comparison of DRFs computed on empirical data (black triangles) and surrogates of the node traits (green crosses), corresponding to the null hypotheses $\mathcal{H}_0^1$ through $\mathcal{H}_0^3$.
	It can be observed that neither A) the preservation of the average trait $O$ ($\mathcal{H}_0^1$), nor B) the additional preservation of each individual node's average trait $O_i$ ($\mathcal{H}_0^2$) is sufficient to reproduce the data. 
	C) However, when the individual node persistence, defined as the inverse of the number of trait switches, is also conserved ($\mathcal{H}_0^1$), the surrogate and empirical data show good agreement. 
	Thus, we do not find sufficient evidence that contagion plays a significant role.
	%\jona{Error bars for the empirical and surrogate DRFs (pooled over all surrogate realisations) are based on a conservative estimate of the standard error of the switching probabilities 
    %\jakob{$\p(K)$}
	%$p(K)$
	%(Eq.~\ref{eq:error-bars}). 
	\niklas{Error bars are computed as described in Sect.~\ref{sec:dose_response_fct}.}
	\jona{Confidence bounds for surrogate DRFs are 
	\jakob{the 95\,\% confidence interval of}
	%based on 
	the distribution of 
	\jakob{$\p(K)$}
	%$p(k)$ 
	over all surrogate realisations.}
	}
	\label{fig:trait_surrogates}
\end{figure}

We test the most basic assumption of whether the empirical DRF can be explained by uncorrelated traits. To do so, all traits were uniformly permuted at random and only the global mean activity level $O = \overline{\ev{o_i(t)}_i}$, was conserved. Here, the overline and the brackets represent the time and ensemble mean, respectively. All possible contagion dynamics are destroyed in the model due to the random permutations.\\
\textbf{Expectation.} We expect to observe no correlation between the DRF $\pas$ of the surrogate and $K$ due to the permutations. Moreover, $\pas(K)$ should be equal to the fraction of active states in the whole observed period.
\\
\textbf{Result.} In Fig.~\ref{fig:trait_surrogates}A, the DRF $\pas$ of the surrogate is contrasted with the empirical DRF $\pa$. We find our expectations confirmed, $\pas$ is quantitatively and qualitatively different from $\pa$. Moreover, $\pas$ is approximately equal to the share of active states. 
\jakob{We quantify the observed difference using the $\zeta$ test statistic introduced in Sec. \ref{sec:surrogate_data}. For the here discussed DRFs the score is $\zeta = 328 \gg 1$.}
Therefore, the model is not sufficient to explain the empirical dynamics and we reject the first null hypothesis.

\paragraph{Second Data Test. Hypothesis $\mathcal{H}_0^2$: $\niklas{P(A_{ij}(t), O_i)}$.}
\textit{The empirical DRF can be reproduced with a class of models that is based only on each node’s individual activity level $O_i=\overline{o_i(t)}$.} 

We test the effects of the individual activity level of each node $O_i = \overline{o_i(t)}$. Analogous to the previous model, the traits per node are randomly permuted in time, but this time only within each node's time series. Therefore, $O_i$ is conserved. As in the previous model, any possible contagion dynamics are destroyed due to the permutations.\\
\textbf{Expectation.} Due to the permutation in the surrogate, the individual probability of the node to change its trait is equal to $O_i$. In particular, this probability is independent of the exposure $K$. Therefore, we do not expect any correlation between $\pas$ and $K$.
\\
\textbf{Result.} Contrary to our expectations, in Fig.~\ref{fig:trait_surrogates}b we find the probability $\pas$ and $K$ positively correlated, qualitatively similar to the correlation of $\pa$ and $K$. However, for $K>100$, the probability $\pas(K)$ continues to increase, while $\pa(K)$ appears to saturate. Furthermore, $\pas$ and $\pa$ differ quantitatively by a factor of about six. Thus, the conservation of $O_i$ is not sufficient to explain the empirical DRF
\jakob{$\zeta = 309 \gg 1$}
, and we also reject the second null hypothesis.

In the second considered model, we found that the DRFs of the surrogate and the empirical data behave in a qualitatively similar way. This could be the result of pre-existing clustering in the data set: contacts $j$ of nodes $i$ would have similar activity values $O_j \approx O_i$ over the entire observation period. A node $i$ with e.g.\ low $O_i$ thus has contacts $j$ with low $O_j$ and therefore receives low exposure $K$. A positive correlation would be the result. Even without fully understanding the cause of the correlation found, it can be concluded that the individual activity level $O_i$ is an essential feature in the empirical network. 
In addition to the correlation, we found a shift of the DRF $\pas(K)$ by a factor of six compared to $\pa$. We suspect the reason for this shift to be the non-preserved persistence of the nodes (inverse number of individual activity state changes). Due to the random permutations, the nodes change their trait more frequently than in the empirical network. In the following surrogate, this hypothesis is analysed in more detail.

\paragraph{Third Data Test. Hypothesis $\mathcal{H}_0^3$: $\niklas{P(A_{ij}(t),\{\tau_{i;0,1}\})}$.}
\textit{The empirical DRF can be reproduced with a class of models that is based only on each node’s individual activity level $O_i$, and its individual persistence (inverse number of individual activity state switches).} 

Additionally to $O_i$, the effect of individual persistence is tested. To achieve this, both the intervals with active trait $o_i(t) = 1$ and the intervals with passive trait $o_i(t) = 0$ were permuted at random. Hence, $O_i$ and the persistence are conserved. Similar to the previous models, the random permutations remove any possible contagion dynamics.\\
\textbf{Expectation.} Due to the additional conservation of individual persistence, we expect $\pas$ to be qualitatively similar to $\pas$ from the second model, but shifted closer to the empirical DRF on the y axis.\\
\textbf{Result.} In Fig.~\ref{fig:trait_surrogates}C, we find, consistently with our expectations, that the DRF of the surrogate is shifted. Moreover, the probability $\pas$ saturates for $K>100$, analogous to the empirical DRF. 
\jakob{Using the $\zeta$ test statistic, no significant deviation $\zeta = 0.79 < 1$}
between $\pas$ and $\pa$ can be found. Therefore, we do not reject the third null hypothesis.\\

The third model showed that individual persistence is a main feature in the empirical network. 
Moreover, the model reproduces the empirical DRF in the model even without contagion.
Thus, the third model shows that the data are not sufficient evidence that contagion plays a significant role in the empirical network, contrary to the hypothesis we formed when we first observed the correlation of $\pa$ and $K$.

\subsubsection{Investigation for Group Dynamics}
\label{sec:investigation_group}

In the previous section, we tested the effects of individual properties such as the individual activity level $O_i$ or the individual persistence with our models. To investigate the importance of group dynamics, in this section we discard all individual properties and test the following null hypothesis:

\begin{figure}[ht]
	\centering
	\includegraphics[width=\Lscale\linewidth]{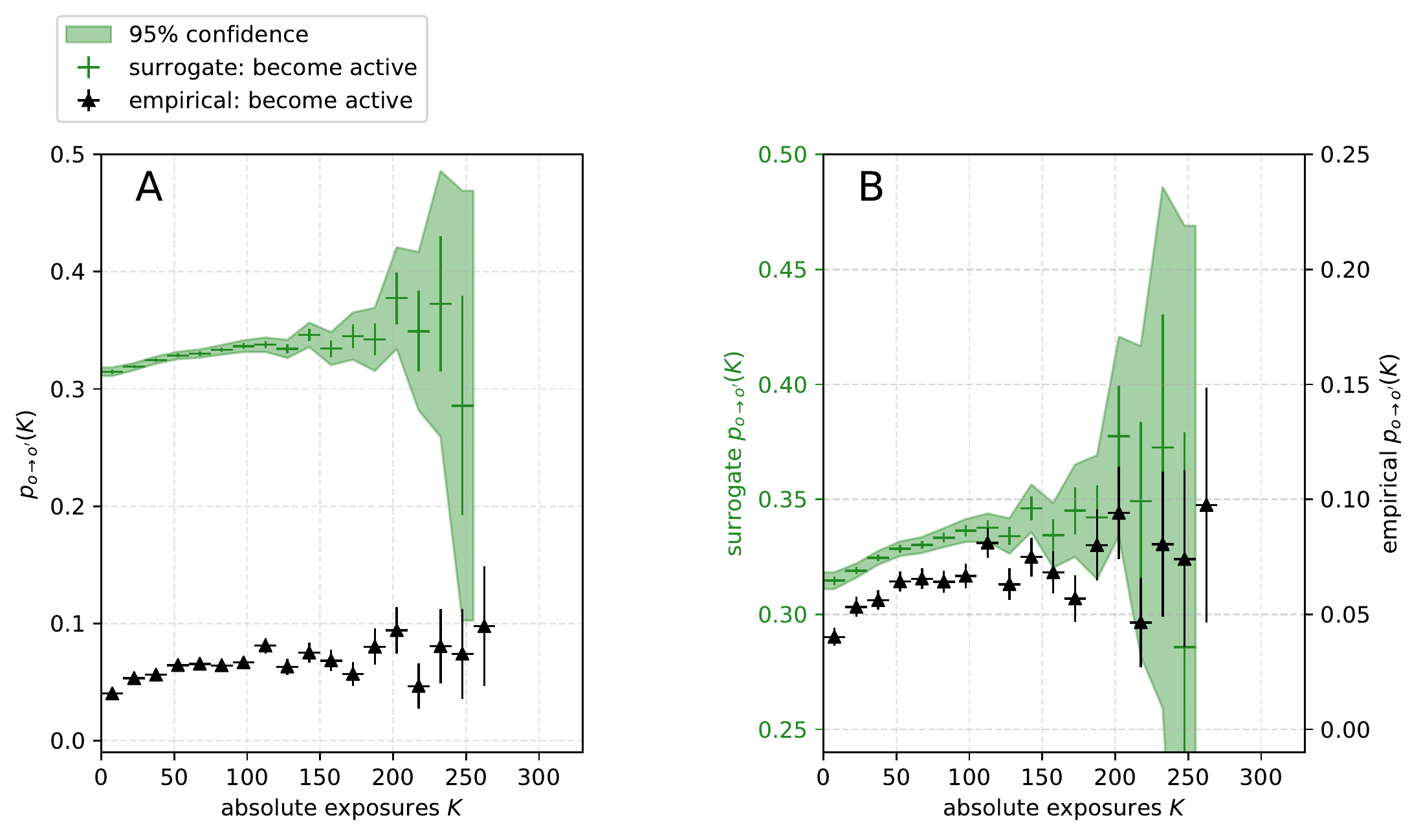}
	\caption{
	Comparison of the DRF for empirical (black triangles) and surrogate (green crosses) data for null hypothesis ($\mathcal{H}_0^4$).
	To investigate external influences that affect all nodes simultaneously, the node traits were randomized in a way that conserves the time-varying mean activity level $O(t)$ of the group.
	The two figures contain the same data: A) compares the absolute values of the data points, while in B) the surrogate data y-axis (green, left side) is offset by 0.25 to facilitate comparison of the functional forms.
	While the absolute values differ strongly, similarities in the functional forms are apparent, pointing to the importance of external influences on the collective group dynamics.
	%\jona{Error bars for the empirical and surrogate DRFs (averaged over all surrogate realizations) are based on a conservative estimate of the standard error of the switching probabilities $p(K)$ (Eq.~\ref{eq:error-bars}). Confidence bounds for surrogate DRFs are based on ... .}
	\niklas{Error bars are computed as described in Sect.~\ref{sec:dose_response_fct}.
	Confidence bounds for surrogate DRFs are the 95\,\% confidence interval of the distribution of $\p(K)$ over all surrogate realisations.}
	}
	\label{fig:trait_surrogate_groupdyn}
\end{figure}

\paragraph{Fourth Data Test. Hypothesis $\mathcal{H}_0^4$: $\niklas{P(A_{ij}(t), O(t))}$.}
\textit{The empirical DRF can be reproduced with a class of models that is based only on the mean time-dependent activity level $O(t)=\langle o_i(t)\rangle_i$ of the ensemble.}

We test the relevance of the mean time-dependent activity level $O(t) = \ev{o_i(t)}_i$ for the empirical dynamics. To do this, the traits between nodes were permuted at random for each time point separately, and only $O(t)$ is preserved.\\
\textbf{Expectation.} Given the permutations, both the probability of becoming active $\pas$ and the exposure $K$ depend on $O(t)$. Thus, a correlation between $\pas$ and $K$ is to be expected.
Furthermore, we expect $\pas(K) \gg \pa(K)$ resulting from the destruction of the persistence of the nodes.
\\
\textbf{Result.} Fig.~\ref{fig:trait_surrogate_groupdyn}A compares the DRF $\pas$ obtained from the surrogate data to the empirical DRF $\pa$. Fig.~\ref{fig:trait_surrogate_groupdyn}b shows the same DRFs, but the DRF of the surrogate (green, left y-axis) is offset by 0.25 to better compare the shape of the functions.
In line with our expectations, $\pas$ is correlated with $K$. For $K<100$, the probability $\pas(K)$ increases linearly. The empirical $\pa(K)$ also increases for $K<100$, but slightly non-linearly. Quantitatively, we observe $\pas(K) \gg \pa(K)$. 
Thus, without individual traits, the model is not able to reproduce the empirical DRF \jakob{$\zeta =326 \gg 1$}. Therefore, we reject the fourth null hypothesis.

Although the surrogate model DRF is quantitatively significantly different from the empirical DRF, the model predicts a qualitatively similar functional form. 
Temporal group dynamics thus seems to be another important feature in the empirical temporal network data. 
Apparently, participants change their behaviour collectively, as is also evident from the fluctuations observed in the mean activity level (Fig.~\ref{fig:cdf}). 
Such non-stationarities could emerge from internal collective dynamics or be due to external influences such as, for example, exam periods, weekends or holidays. 
A more detailed analysis is needed to distinguish these possible effects.

\subsubsection{Investigation for Homophily Dynamics}
\label{sec:investigation_homophily}

Continuing our investigation, we look for homophily dynamics in the network. 
Analogously to the analysis testing for contagion effects, we create surrogate models in which explicitly no homophily takes place. 
With these, we attempt to reproduce the empirical dynamics.
To this end, we permute the network edges $A_{ij}(t)$ and keep the properties of the nodes $o_i(t)$ unchanged. 
This approach removes any homophily dynamics from the network, since the drawing and breaking of edges is randomised. 
The investigation is carried out in two steps, testing the following null hypotheses:

\paragraph{Fifth Data Test. Hypothesis $\mathcal{H}_0^5$: $\niklas{P(A, O_i(t))}$.} 
\textit{The empirical DRF can be reproduced with a class of models that is based only on individual activity dynamics and the average network edge density $A=\overline{\langle A_{ij}(t)\rangle_{i,j}}$.} 

\begin{figure}[h]
	\centering
	\includegraphics[width=\Lscale\linewidth]{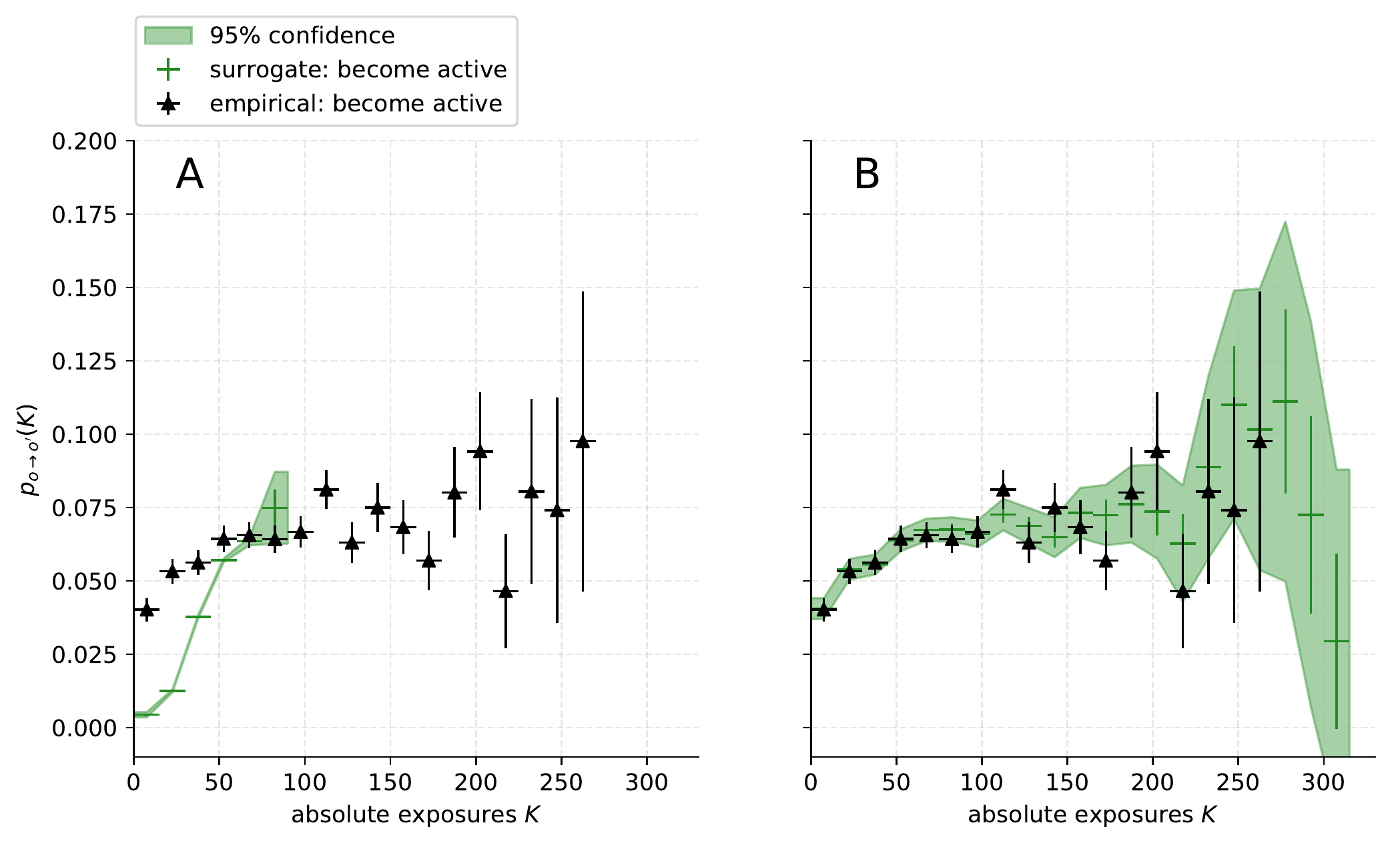}
	\caption{
	Comparison of DRFs computed on empirical data (black triangles) and surrogates of the network topology (green crosses) for null hypotheses ($\mathcal{H}_0^5$) and ($\mathcal{H}_0^6$).
	In A) only the mean node degree $k$ is conserved ($\mathcal{H}_0^5$), leading to a significant difference between empirical and surrogate data.
	In B) each node's time-varying degree $k_i(t)$ is conserved as well ($\mathcal{H}_0^6$), corresponding to a test for homophily in the network, with good agreement between the DRFs.
	It can be concluded that, while the non-trivial network structure appears to be of importance, no significant evidence for homophilic dynamics can be found.
	%\jona{Error bars for the empirical and surrogate DRFs (averaged over all surrogate realisations) are based on a conservative estimate of the standard error of the switching probabilities $p(K)$ (Eq.~\ref{eq:error-bars}). Confidence bounds for surrogate DRFs are based on ... .}
	\niklas{Error bars are computed as described in Sect.~\ref{sec:dose_response_fct}.
	Confidence bounds for surrogate DRFs are the 95\,\% confidence interval of the distribution of $\p(K)$ over all surrogate realisations.}
	}
	\label{fig:edge_surrogates}
\end{figure}

We test the most basic assumption that the empirical dynamics can be explained by a random network. 
For this purpose, all edges were permuted uniformly at random. 
Only the average temporal network edge density $A=\overline{\langle A_{ij}(t)\rangle_{i,j}}$ was conserved. 
In this model, any homophily dynamics is removed, as the formation and breaking of edges is randomized.
\\
\textbf{Expectation.} Since the traits have been kept unchanged, we expect the DRF of the model and the empirical DRF to be of the same order of magnitude. 
Due to the randomisation of the network, the neighbourhoods of the nodes are randomised as well. 
Thus, no correlation between the exposure $K$ received from the neighbours and the probability $\pas$ of changing the trait is to be expected.
\\
\textbf{Result.} The DRF of the model and the empirical DRF are compared in the Fig.~\ref{fig:edge_surrogates}A. 
Contrary to our expectation, we can observe a correlation between $\pas$ and $K$. 
Moreover, for the model, the case $\pas(K)$ for $K>100$ does not exist.
Both DRFs have the same order of magnitude, which is in line with our expectations. 
However, only a few bins of the empirical DRF lie within the 95\% confidence interval of the DRF from the surrogate
\jakob{and calculating the $\zeta$ test statistic gives $\zeta = 61 > 1$.}
Consequently, we reject the fifth null hypothesis.

When analysing our model based on a random network, we observed a positive correlation between $\pas$ and $K$. 
This correlation was significantly different from the correlation found for the empirical DRF. 
Therefore, the non-trivial network structure and dynamics appear to be essential for reproducing the empirical dynamics.
One explanation for the correlation found could be the external influences already described in Sect.~\ref{sec:investigation_group}.  Nodes may change their traits in synchrony, independently of the network and caused by an external influence. 
This would affect $K$ as well and could explain the correlation found. 
A further analysis is necessary here.
Another feature of the surrogate model's DRF is that no large exposure $K>100$ occurred. 
This is likely caused by a much smaller variance of the degree distribution in the random network than in the empirical one.
In the following surrogate, this hypothesis is analysed in more detail.

\paragraph{Sixth Data Test. Hypothesis $\mathcal{H}_0^6$: $\niklas{P(k_i(t), O_i(t))}$.} 
\textit{The empirical DRF can be reproduced with a class of models that is based only on the individual node dynamics, and each node's time-dependent network degree $k_i(t)=\sum_{j=0}^N A_{ij}(t)$.} 

Building on the previous model 
we test whether the time-dependent network degree of the nodes $k_i(t)=\sum_{j=0}^N A_{ij}(t)$ has a significant impact on the network dynamics. 
For this purpose, the edges of the network are permuted at random, but $k_i(t)$ is preserved. 
Analogous to the previous model, the homophily dynamics is removed by the permutations.\\
\textbf{Expectation.} For the correlation of $\pas$ and $K$ we expect it to be similar to the one of the previous model. 
However, for this model we conserved the node's degree. 
Thus, the progression of the DRF should also extend over $K>100$.\\
\textbf{Result.} In Fig.~\ref{fig:edge_surrogates}b we compare the DRF of the model with the empirical one. In agreement with our expectation, we find $\pas(K)$ for $K>100$. 
However, the correlation of $\pas$ and $K$ is different from the previous model (Fig.~\ref{fig:edge_surrogates}A). 
No significant difference \jakob{$\zeta = 0.31 < 1$} to the empirical DRF can be found anymore, \jakob{using the $\zeta$ test statistic}.
Therefore, we cannot reject the sixth null hypothesis.\\

With this final surrogate model, we were able to reproduce the empirical DRF by conserving the node degree sequence in the temporal network data. 
Accordingly,  node degree $k_i(t)$, the number of social contacts a student has at a given time $t$ within the student population covered by the study, seems to be an important feature in the empirical data set. 
Furthermore, the reproduction succeeded without including the dynamics of homophily. 
%This shows that the empirical data provide not only no sufficient evidence for a significant influence of contagion (see the results for $\mathcal{H}_0^3$ reported above), but are also not sufficient evidence for a significant influence of homophily either. 
% Note: This was criticised by a reviewer for being too definite; the data could contain evidence that our method can't detect. I suggest this instead: 
\niklas{Thus, we do not detect a significant influence of contagion (see the results for $\mathcal{H}_0^3$ reported above), but neither a significant influence of homophily.}

\subsection{Summary}\label{sec:summary}

\begin{figure}[h]
	\centering
	\begin{subfigure}[b]{0.49\textwidth}
         \centering
          \includegraphics[width=\textwidth]{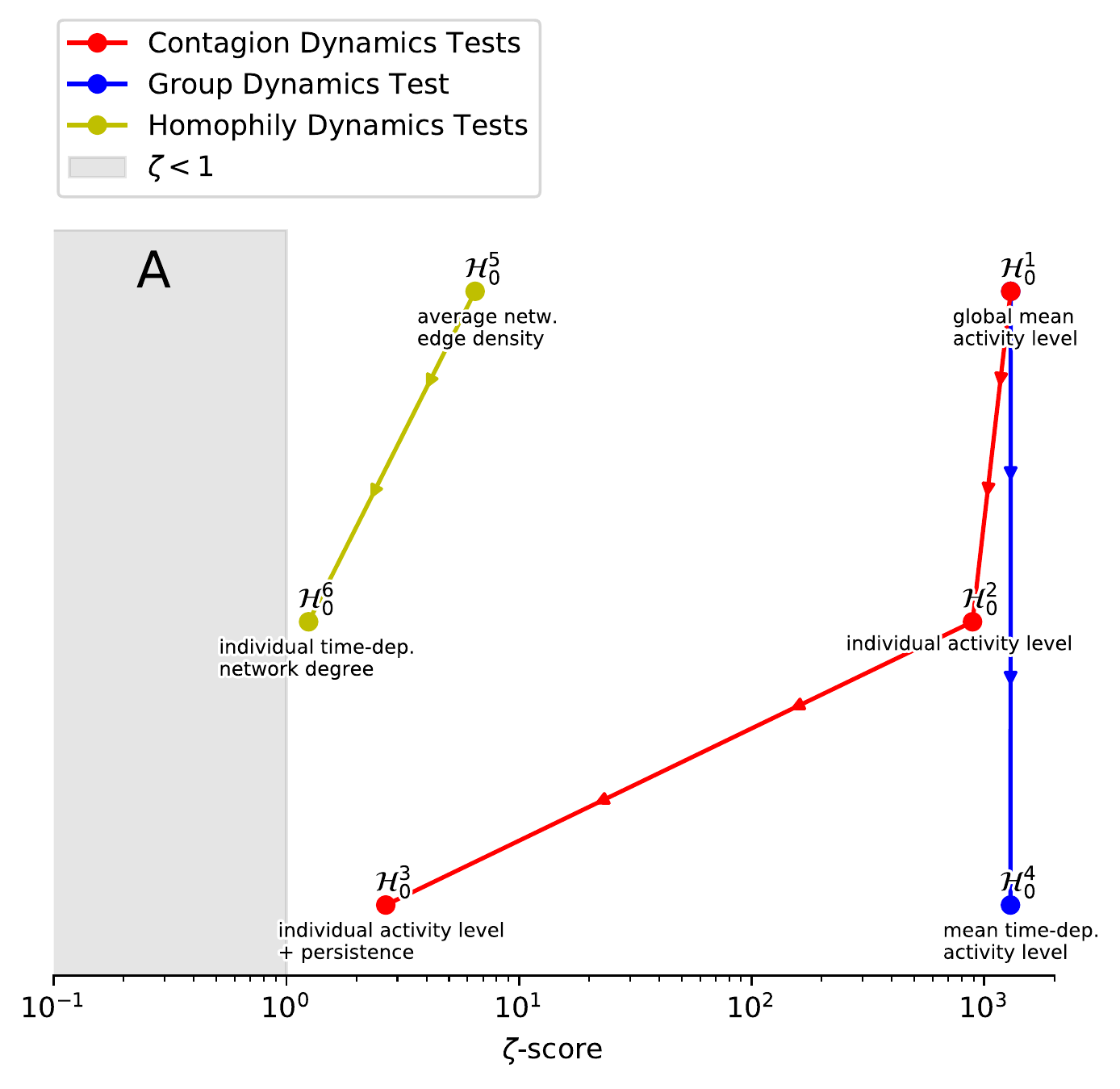}
          \caption{$\zeta$-scores for the AVM data ($\varphi = 0.6$).}
     \end{subfigure}
	\begin{subfigure}[b]{0.49\textwidth}
         \centering
         \includegraphics[width=\textwidth]{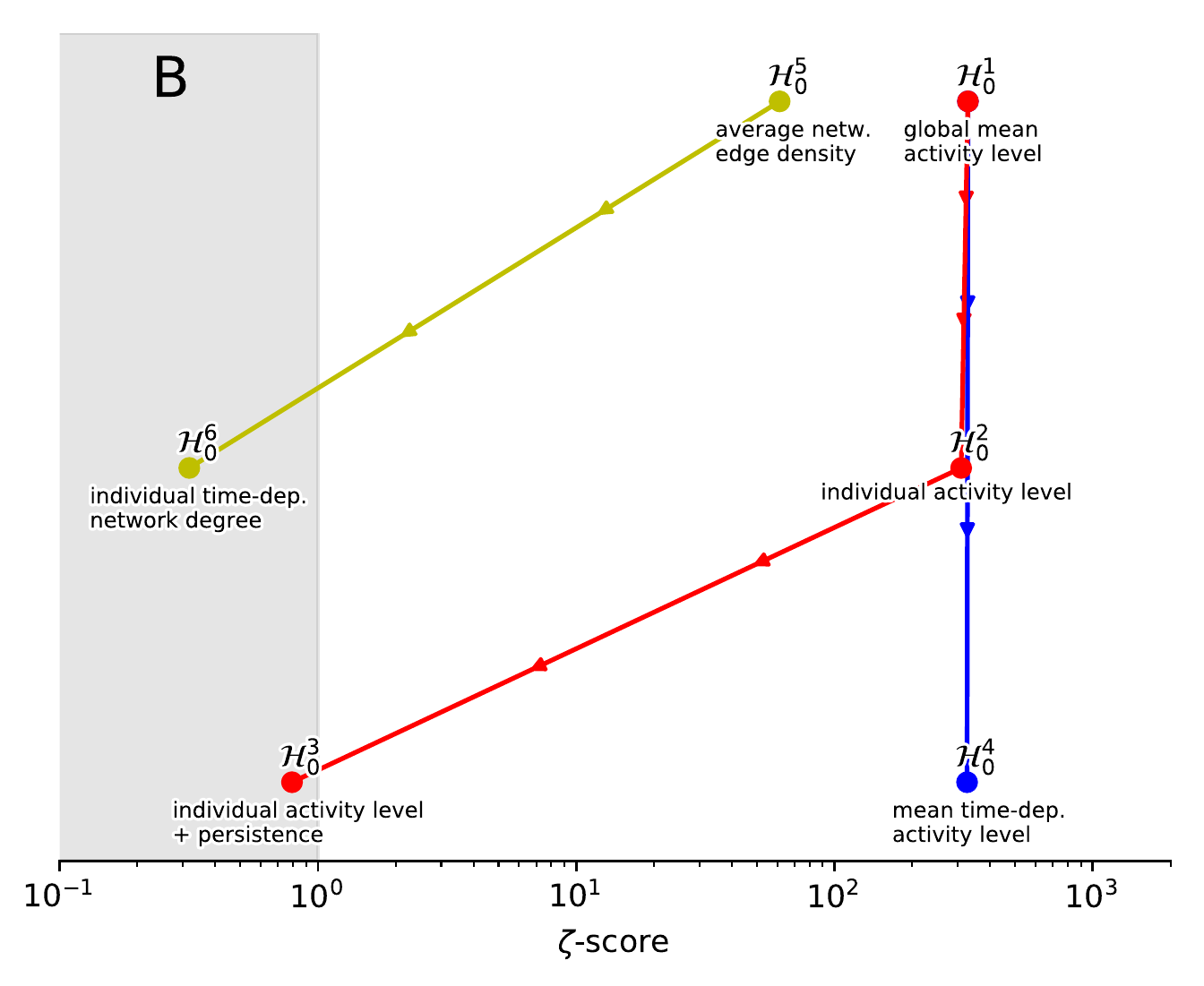}         
         \caption{$\zeta$-scores for the CNS data.}
     \end{subfigure}
     \hfill
	\caption{\jakob{Comparison of $\zeta$-scores for a hierarchy of surrogate tests, for (A) the synthetic AVM data ($\varphi = 0.6$) and (B) the empirical CNS data. Each circle in the figure represents a single surrogate data test. The horizontal location of the circle reports the $\zeta$-score of the tested hypothesis. An arrow from a surrogate test at a higher location to a lower one indicates that the former randomises more structure in the data than the latter. The null hypothesis name of each test is given above each circle, and the conserved features of the surrogate model below it (see Sect. \ref{sec:surrogate_data}). The link and circle colour indicate which dynamics were investigated with the tests. The red and the blue branches give the class of surrogate tests with permuted traits $o_i(t)$, while for the yellow branches the edges $A_{ij}$ were permuted at random. The grey rectangle marks the area where the empirical DRF does not differ significantly from the surrogate DRF.}}
	\label{fig:surrogate_overview}
\end{figure}

\jakob{In the sections \ref{sec:results_avm} and \ref{sec:results_empirical} we presented the results of our methodology, which we applied first to synthetic data from the Adaptive Voter Model (AVM) and secondly to empirical data from the Copenhagen Networks Study (CNS). 
For both the synthetic and the empirical DRF, we found a monotonic functional dependency. 
In the synthetic case, it arises from the dynamics of the model: homophilic rewiring and social learning.
To investigate whether contagion and homophily are the main driver for the empirical DRF, six null hypotheses $\mathcal{H}_0^1$ to $\mathcal{H}_0^6$ were tested.
The tests were conducted by analysing two classes of surrogate models. 
In one, the traits $o_i(t)$ and in another, the edges $A_{ij}(t)$ were randomly permuted. 
Each class consists of a hierarchy of surrogate models.
Starting with the most basic  \marc{model}, %where
\marc{in which} all traits resp. edges are randomly permuted, we gradually conserve parts of the system until the surrogate DRF $\pas(K)$ and the empirical DRF $\pa(K)$ are \marc{considered} equal within an error margin.
As proof of concept, this methodology was applied 
%on
to the synthetic DRF \marc{of an adaptive voter model} (see Appendix \ref{apx:AVMsurrogates} for detailed results).
In Fig.~\ref{fig:surrogate_overview} we present a result-compilation of the test hierarchy for the synthetic data (A) of the AVM ($\varphi = 0.6$) as well as for the empirical data (B).
The red and the blue branches give the class of surrogate tests with permuted traits $o_i(t)$, while for the yellow branches the edges $A_{ij}$ were permuted at random. An arrow from a surrogate test at a higher location to a lower one indicates that the former shuffles more than the latter. 
The differences between $\pas(K)$ and $\pa(K)$ are %plotted 
\marc{displayed}
on the horizontal axis and was quantified using a test statistic $\zeta$ introduced in Sec. \ref{sec:surrogate_data}.
For the synthetic data (A) the yellow and the red branch end with $\mathcal{H}_0^3$ and $\mathcal{H}_0^6$ outside the grey area ($\zeta \geq 1$), indicating a significant difference between $\pas(K)$ and $\pa(K)$.
Since we test with $H_0^3$ ($H_0^6$) whether the DRF can be explained without contagion (homophily), but both are core dynamics in the underlying model, this result was expected.
In contrast, for the empirical data (B) $\mathcal{H}_0^3$ and $\mathcal{H}_0^6$ lie within the grey area, indicating no significant difference between $\pas(K)$ and $\pa(K)$.
Consequently, this leads to the conclusion that we find neither significant evidence for an influence of contagion nor significant evidence for homophily in the CNS data.
Considering all the tests performed on the empirical data, individual activity level, individual behavioural persistence, the effects of a possibly externally forced collective group dynamic and the individual number of social contacts (the node degree sequence) are sufficient to explain the estimated empirical DRF.
}

\section{Discussion and Conclusion}
\label{sec:disc_concl}
In this paper, we proposed a methodology for estimating dose response functions (DRFs) from temporal network data. 
We developed a hierarchy of surrogate data models to evaluate to what degree the observed DRFs can be explained by underlying processes such as social contagion, collective group dynamics and homophily. 
These surrogate models test the effects of distinct data features, such as overall and individual node activity levels, individual node trait persistence, overall network link density and individual node degrees. 
We applied this methodology to empirical temporal network data from the Copenhagen Networks Study, focusing on the illustrative health-related behaviour ``regularly going to the fitness studio'' in a physically-close-contact network of 619 %850
 university students, observed over the course of three months. 
\niklas{We find }%The empirical data 
neither 
%provide 
significant evidence for an influence of contagion, nor significant evidence for homophily.
The individual activity level, individual behavioural persistence, effects of possibly externally forced collective group dynamics, and individual number of social contacts (the node degree sequence) are sufficient to explain the estimated empirical dose response function. 
\niklas{These %negative 
findings are underlined by a validation study performed using synthetic data, in which the sensitivity of our methodology to contagion and homophilic effects is demonstrated.}

In the context of the application case considered in %our
\marc{the present} study, %these
\marc{our} findings contradict the perspective that social interactions influence adopted behaviour, for example via subjective norms \cite{ajzen_theory_1991}, as supported by psychological research \cite{bandura_social_1999}. 
In particular, the ability of social norms to influence individual decision-making has been identified previously as a potential tool for large-scale group behaviour transformations \cite{nyborg2016social, young_evolution_2015}.
However, in the present context of exercise behaviour a person may only be susceptible to social influence during particular stages of their decision process, while being almost ``immune’’ at other times \cite{prochaska_transtheoretical_1994, marcus_transtheoretical_1994}. 
At any time, too few people may be in this socially susceptible state to rise above the noise threshold in the data.

Overall, our results demonstrate that care needs to be taken in interpreting dose response functions obtained from empirical temporal network data;
in particular when considering observational data that did not emerge from experiments in more controlled environments \cite{bond201261,kramer2014experimental}. 
Even pronounced positive correlations between exposure to a trait and the probability to adopt this trait can arise from structures in the temporal network data that do not need to be related to contagion and spreading processes, or homophily. 
Applying and further developing methodologies based on hierarchies of surrogate models, such as the one proposed in this article, provides a way forward to discern the specific imprints of complex spreading processes in temporal network data. 
Cases where the presence of such processes is not supported by the data can thus be excluded.

Our analysis has limitations in several dimensions that should be considered. 
Firstly, in terms of data limitations, the empirical temporal network data set extracted from the Copenhagen Networks Study depends on multiple assumptions on thresholds and other parameter values. 
The definition of social contacts as links in a physically-close-contact network could be too unspecific for discerning social contagion effects. 
Social contagion might be expected to require a more permanent and intense social relationship such as friendship to be effective. 
\niklas{Likewise, the chosen 1-day timescale of the contact network %social contact 
may need to be reconsidered, as clustering in the CNS data has been shown to disappear at time scales greater than one hour \cite{sekara2016fundamental}.}
Furthermore, the definition of node traits as active or passive may suffer from noise and missing data issues, since
most likely some fitness studios and other relevant exercise institutions (e.g. university gyms, swimming pools etc.) are missing from our list. 
Also, using GPS coordinates to determine whether a student is visiting a fitness studio introduces uncertainties: in a densely populated urban area like the city of Copenhagen, a café or a library might be located right next to, or even above or below a fitness studio, introducing additional noise into our data set.

Secondly, considering methodological limitations, DRFs are a highly aggregate statistical indicator describing a complex temporal network data set.
They might not be specific enough to detect subtle spreading processes or to discriminate different types of complex contagions.
Arguably this calls for higher order statistics with larger statistical power. 
Moreover, the proposed methodology based on a hierarchy of surrogate data sets is limited in that it allows only for indirect inference on the possible presence of spreading or contagion processes.
In this respect it is desirable to augment the present analysis with more direct investigations including generative models of complex network spreading processes.

In summary, we suggest that our methodology is promising for applications to other systems and temporal network data sets.
This can, among other applications, possibly aid our understanding of the social dynamics, spreading potentials and possible social tipping points in behaviours and social norms relevant for the adoption of healthy and sustainable diets \cite{willett2019food}
that can help to feed the world within planetary boundaries \cite{gerten2020feeding}. 
Efforts should be directed towards providing high-quality empirical temporal network data sets that can be leveraged for understanding complex spreading processes in these relevant domains. 
Promising directions of methodological developments include higher order statistics such as multi-node correlations for discerning the effects of longer contagion chains, spreading contagion waves, or the imprints of network motifs on complex spreading processes. 
Astute surrogate data models can provide detailed insights into such spreading processes. 
Connecting empirical network data to generative statistical and dynamical adaptive network models more directly, e.g.\ via maximum likelihood methods, appears similarly promising. 
Hence, one can open new perspectives to predict future spreading dynamics. 
Ultimately, this research thus aids in designing targeted interventions for fostering desirable or suppressing unwanted contagions in diverse complex systems including pandemics, \niklas{the} brain, traffic and sustainability transformations.

\section*{Acknowledgements}
The authors would like to thank Franziska Gutmann and Michaela Schinkoeth of the Sport and Exercise Psychology research group at University of Potsdam for a helpful discussion.
JFD, JH, JHL and MW are thankful for financial support by the Leibniz Association (project DominoES). 
JFD acknowledges support from the European Research Council project Earth Resilience in the Anthropocene (743080 ERA).
NK is grateful to the Geo.X Young Academy for financial support.
SL acknowledges support by the Danish Research Council and the Villum Foundation.

\section*{Author contributions}

\niklas{
SL conducted the Copenhagen Network Study, and provided the experimental data.
JFD, JH, JV conceived this study.
JHL curated the data, implemented most of the methods and simulations, and visualised the results, with supervision by JFD and JV.
NHK implemented the AVM-based method validation, with support from JHL.
All authors interpreted and discussed the results.
JFD, JHL, and NHK wrote the manuscript, with support from JH, SL, JV and MW.
All authors give their final approval of the article version to be published.
}
 
\bibliographystyle{naturemag}
\bibliography{bib} 

% \begin{thebibliography}{}
% % and use \bibitem to create references.
% \bibitem{RefJ}
% % Format for Journal Reference
% Author, Journal \textbf{Volume}, (year) page numbers
% % Format for books
% \bibitem{RefB}
% Author, \textit{Book title} (Publisher, place year) page numbers
% % etc
% \bibitem{Sekara2014}
% V. Sekara and S. Lehmann, PLoS One \textbf{9}, (2014) 1–8
% \end{thebibliography}
\newpage
\appendix

\niklas{
\section{Surrogate method validation with synthetic data}\label{apx:AVMsurrogates}
}
To evaluate how well the surrogate model method performs, we apply it to the two synthetic data sets created with %``realistic''
\niklas{CNS-aligned} parameters for Fig.~\ref{fig:DRF_example}A.
This data set is generated using the Adaptive Voter Model (AVM), once with ($\varphi=0.6$) and once without ($\varphi=0.0$) the network adaptation process.
Other model parameters are chosen to align with the filtered data extracted from the Copenhagen Network Study (CNS): 
the number of nodes $N=619$, the average degree 
\jakob{$\bar k_i = 19$}
%$\ev{k_i} = 13.5$ 
and the number of simulated time steps $\tau=90$. 
The number of model updates per time step is determined empirically, to align the average number of behaviour switches per time step across the entire system with the value found in the CNS data ($40.24\pm0.96$ behaviour switches per time step).
To maintain the comparability to the CNS data, a single simulation run of the AVM model is used, based on which ten surrogate model realisations are computed.

Using AVM-generated data to test the surrogate methods is a natural choice; when compared with e.g. SI(R) models, the AVM can best describe the processes and conditions of the system. For example, the behaviour is already rather common in the population; there is no ``patient zero.'' 
Furthermore, we assume that contact with "infected" (high activity level) individuals may increase infection probability -- but also vice versa, that contact with "uninfected" (low activity level) individuals makes ``recovery'' more likely. 
However, even when aligning the model parameters to the CNS data, it should be noted that the AVM model does not necessarily represent a ``best guess'' for the real-world dynamics, but only an over-simplified stand-in.

\begin{figure}[h]
	\centering
	\includegraphics[width=\Lscale\linewidth]{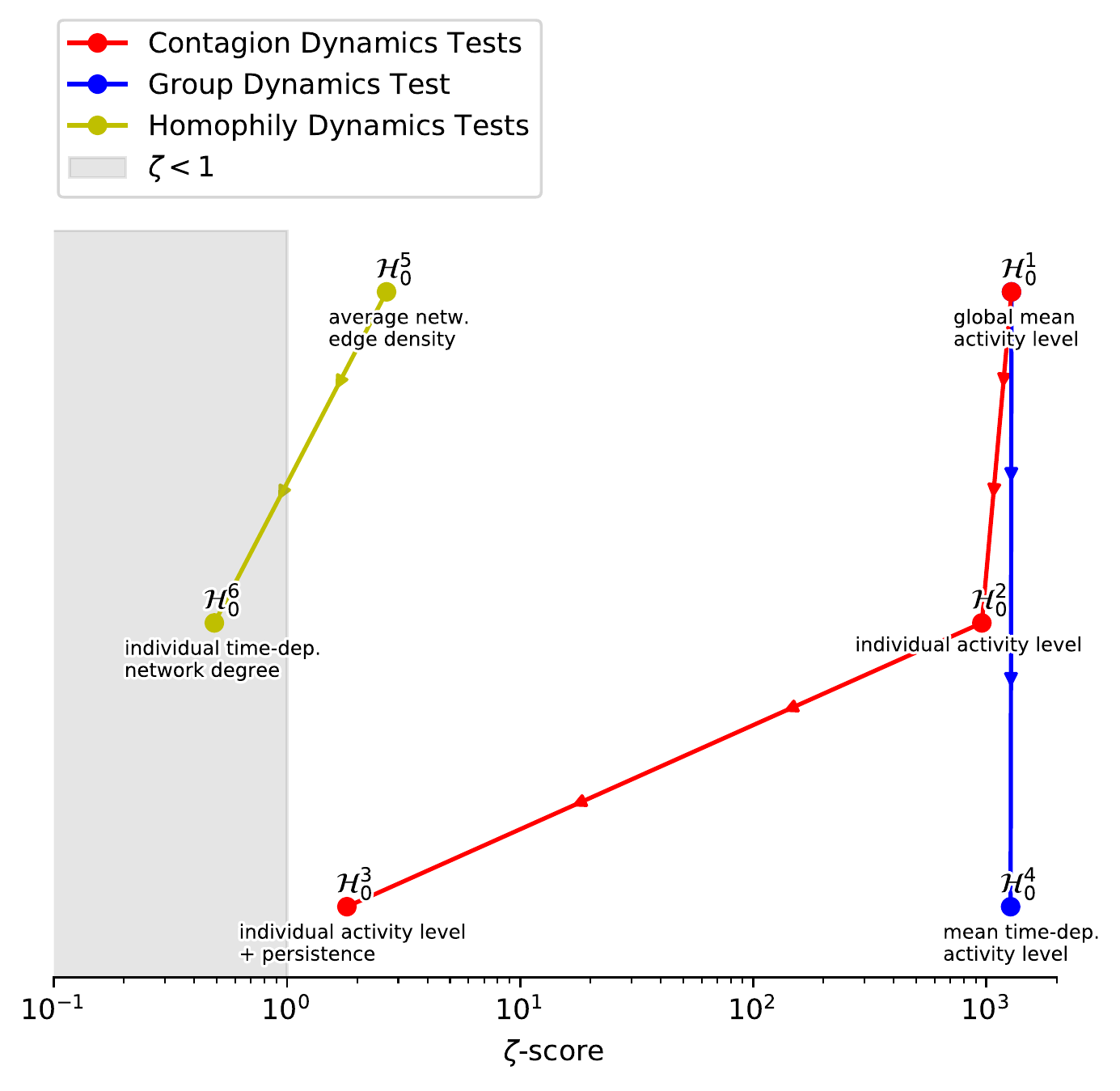}
	\caption{Comparison of $\zeta$-score for the hierarchy of surrogate tests on AVM data, 
	\jakob{for the $\varphi=0.0$ case}.
	%for (A) the $\varphi=0.0$ case and (B) the $\varphi=0.6$ case.
	Each circle in the figure represents a single surrogate data test. 
    \jakob{The horizontal location of the circle reports the $\zeta$-score of the tested hypothesis.}
	%The horizontal location of the circle reports the z-score of the tested hypothesis, given as the combined z-scores from each bin, calculated using Stouffer's method. 
	An arrow from a surrogate test at a higher location to a lower one means that the former shuffles more than the latter. 
	The null hypothesis name of each test is given above each circle, and the conserved features of the surrogate model below it (see Sect. \ref{sec:surrogate_data}). The link and circle colour indicate which dynamics were investigated with the tests. The grey rectangle marks the area where the empirical DRF does not differ significantly from the surrogate DRF.
	}
	\label{fig:surrogate_overview_avm}
\end{figure}

\paragraph{}In the following, we create the hierarchy of surrogate models, which is described in detail in Sect.~\ref{sec:surrogate_data}. In this chapter, ``AVM data'' refers to the synthetic data set generated by the Adaptive Voter Model with the parameters described above, and ``surrogate data'' refers to surrogate data sets created using the AVM data.
\jakob{To quantify the difference between surrogate DRF and AVM DRF we calculate the $\zeta$-score (Eq. \ref{eq:zeta_score}).
A graphical presentation of the test hierarchy can be found in Fig.~\ref{fig:surrogate_overview_avm} for $\varphi = 0.0$, while the corresponding figure for $\varphi = 0.6$ is presented in Sec.~\ref{sec:summary}.}

\begin{figure}[h]
	\centering
	\includegraphics[width=\Lscale\linewidth]{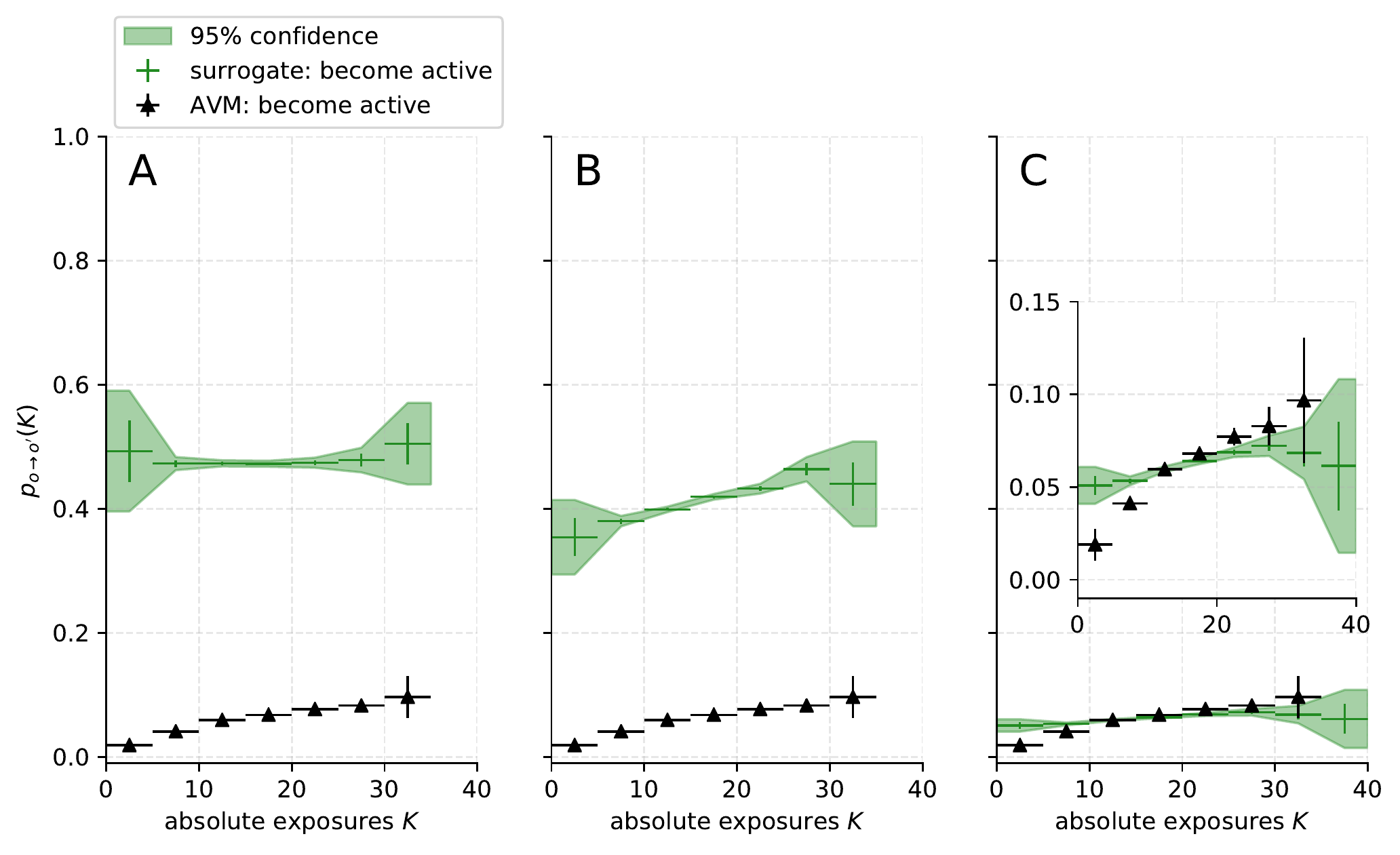}
	\includegraphics[width=\Lscale\linewidth]{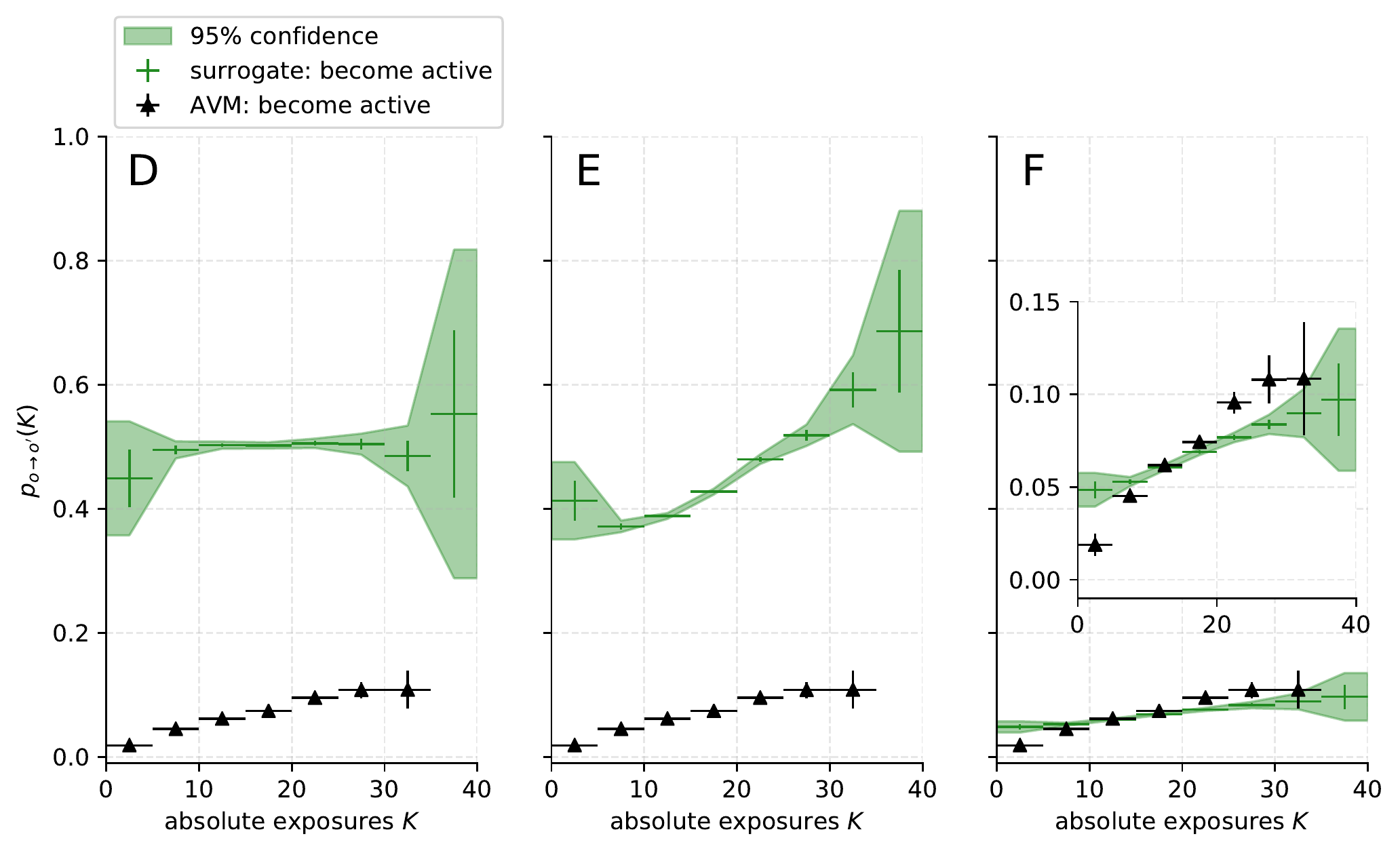}
	\caption{Comparison of DRFs computed on AVM data (black triangles) and surrogates of the node traits (green crosses), corresponding to the null hypotheses $\mathcal{H}_0^1$ through $\mathcal{H}_0^3$.
	(A-C) are for the $\varphi=0.0$ case, while (D-F) correspond to the $\varphi=0.6$ case.
	%\jona{Error bars for the model (averaged over all model simulations???) and surrogate DRFs (averaged over all surrogate realizations) are based on a conservative estimate of the standard error of the switching probabilities $p(K)$ (Eq.~\ref{eq:error-bars}). Confidence bounds for surrogate DRFs are based on ... .}
	\niklas{Error bars are computed as described in Sect.~\ref{sec:dose_response_fct}.
	Confidence bounds for surrogate DRFs are the 95\,\% confidence interval of the distribution of $\p(K)$ over all surrogate realisations.}
	}
	\label{fig:trait_surrogates_avm}
\end{figure}

\paragraph{First AVM Test. Hypothesis $\mathcal{H}_0^1 =P(A_{ij}(t), O)$:} Displayed in Fig.~\ref{fig:trait_surrogates_avm}A and D for $\varphi=0.0$ and $\varphi=0.6$, respectively. 
As could be expected in this complete randomisation of activity states, the DRF becomes flat in both cases, at a level corresponding to the fraction of active nodes in the network.
\jakob{The $\zeta$-score for the run with $\varphi = 0.0$ is $\zeta = 1281$ and the score for $\varphi = 0.6$ is $\zeta=1299$.}

\paragraph{Second AVM Test. Hypothesis $\mathcal{H}_0^2  P(A_{ij}(t), O_i)$:} Displayed in Fig.~\ref{fig:trait_surrogates_avm}B and E for $\varphi=0.0$ and $\varphi=0.6$, respectively.
In this randomisation that conserves the individual node's activity levels, the surrogate DRF is still much higher than the AVM DRF. 
A likely explanation for the rising trends in the surrogate DRFs is the formation of network regions that have relatively homogeneous activity levels through the AVM process.
Such regions, which consist of nodes that lean towards one activity level and whose neighbourhood comprise a majority of nodes with the same activity level, are not destroyed by the $\mathcal{H}_0^2$ shuffling.
This effect can be expected to be stronger for the $\varphi=0.6$ case, where homophilic rewiring is an additional driver in the formation of such regions.
The greater slope in Fig.~\ref{fig:trait_surrogates_avm}E supports this.
\jakob{The $\zeta$-score for the run with $\varphi = 0.0$ is $\zeta = 955$ and the score for $\varphi = 0.6$ is $\zeta=890$.}

\paragraph{Third AVM Test. Hypothesis $\mathcal{H}_0^3  =P(A_{ij}(t),\{\tau_{i;0,1}\})$:} Displayed in Fig.~\ref{fig:trait_surrogates_avm}C and F for $\varphi=0.0$ and $\varphi=0.6$, respectively.
As expected, when conserving the number of behaviour switches, the average switch\marc{ing} probability displayed in the DRF is very similar for the AVM and surrogate data.
However, clear differences between the AVM and surrogate DRFs can be discerned.
\jakob{The $\zeta$-score for the run with $\varphi = 0.0$ is $\zeta = 1.8$ and the score for $\varphi = 0.6$ is $\zeta=2.7$, indicating a significant difference $\zeta > 1$.}
The upward trend of the AVM data DRFs is significantly greater than in the surrogate in both the $\varphi=0.0$ and $\varphi=0.6$ cases.
This is consistent with the true contagion process underlying the AVM simulation data.
This shows the method to be sensitive to contagion effects, implying that the inability to reject $\mathcal{H}_0^3$ in the empirical data (see Fig.~\ref{fig:trait_surrogates}C) is likely due to a lack of dominant contagion dynamics in the studied behaviour.
It should be noted that the surrogate DRFs do not become completely flat, but retain a more moderate upward trend.
This can be explained analogously to the upward trend in the surrogate DRFs of the second AVM test, described above.

\begin{figure}[h]
	\centering
	\includegraphics[width=\Lscale\linewidth]{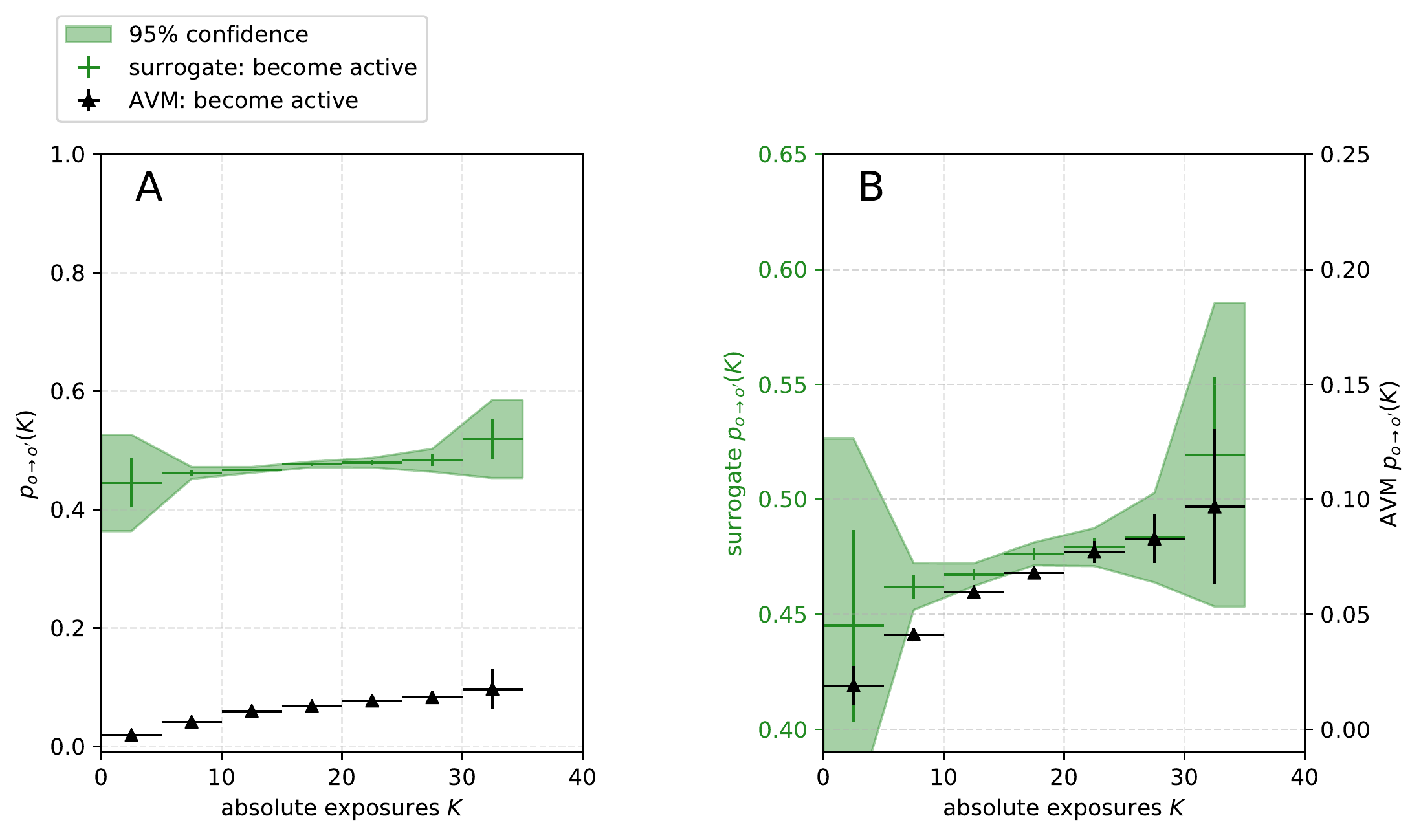}
	\includegraphics[width=\Lscale\linewidth]{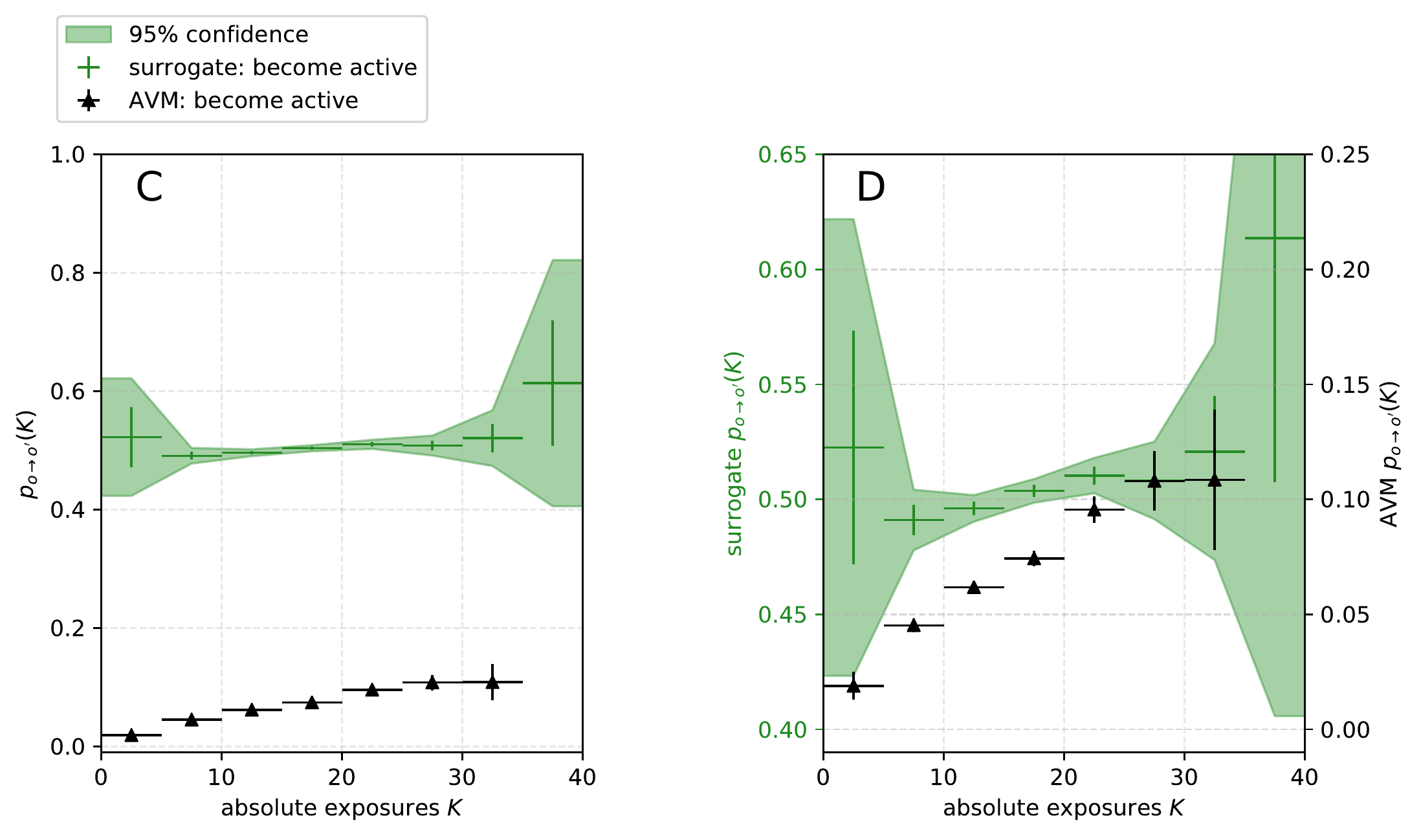}
	\caption{Comparison of DRFs computed on AVM data (black triangles) and surrogates of the node traits (green crosses), corresponding to the null hypothesis $\mathcal{H}_0^4$.
	(A) and (B) are for the $\varphi=0.0$ case, while (C) and (D) correspond to the $\varphi=0.6$ case.
	(B) and (D) display the same data as (A) and (C), respectively, but with \marc{the range of the horizontal} %y-
	axes independently shifted for AVM and surrogate data, to facilitate the direct comparison of the DRFs.
	%\jona{Error bars for the model (averaged over all model simulations???) and surrogate DRFs (averaged over all surrogate realizations) are based on a conservative estimate of the standard error of the switching probabilities $p(K)$ (Eq.~\ref{eq:error-bars}). Confidence bounds for surrogate DRFs are based on ... .}
	\niklas{Error bars are computed as described in Sect.~\ref{sec:dose_response_fct}.
	Confidence bounds for surrogate DRFs are the 95\,\% confidence interval of the distribution of $\p(K)$ over all surrogate realisations.}
	}
	\label{fig:trait_surrogates_groupdyn_avm}
\end{figure}

\paragraph{Fourth AVM Test. Hypothesis $\mathcal{H}_0^4$: $P(A_{ij}(t), O(t))$:} Displayed in Fig.~\ref{fig:trait_surrogates_groupdyn_avm} (A,B) and (C,D) for $\varphi=0.0$ and $\varphi=0.6$, respectively.
The surrogate and AVM DRFs have greatly differing y-scales. 
\jakob{The $\zeta$-score for the run with $\varphi = 0.0$ is $\zeta = 1269$ and the score for $\varphi = 0.6$ is $\zeta=1295$.}
However, in the $\varphi=0.0$ case, the surrogate DRF retains an upward trend, albeit smaller than the AVM DRF. 
Since $\mathcal{H}_0^4$ is essentially the mean-field approximation of the system, this demonstrate\marc{s} how the network is densely, and relatively homogeneously connected in this case.
In the $\varphi=0.6$ case, the randomisation destroys any significant slope.
Here, the original AVM data apparently differs more strongly from the mean-field approximation, which can be explained by the greater degree of homophilic clustering in this case.
The network structure, with its additional rewiring mechanism, thus appears more important in this case.
The behaviour seen in the evaluation of $\mathcal{H}_0^4$ in the empirical CNS data (Fig.~\ref{fig:trait_surrogate_groupdyn}) resembles the $\varphi=0.0$ case in AVM data, which can be interpreted as an absence of clustering in the CNS data.

\begin{figure}[h]
	\centering
	\includegraphics[width=\Lscale\linewidth]{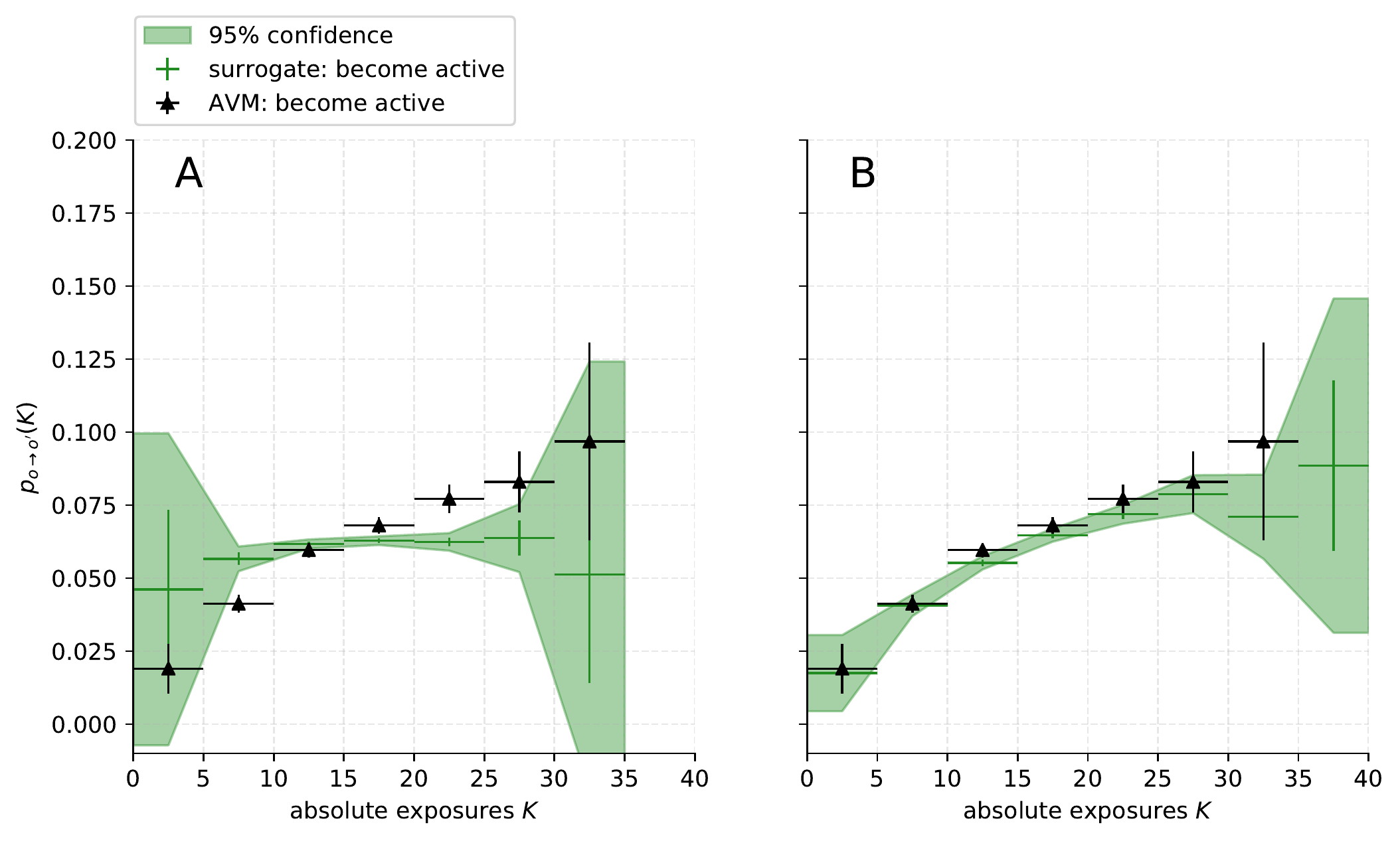}
	\includegraphics[width=\Lscale\linewidth]{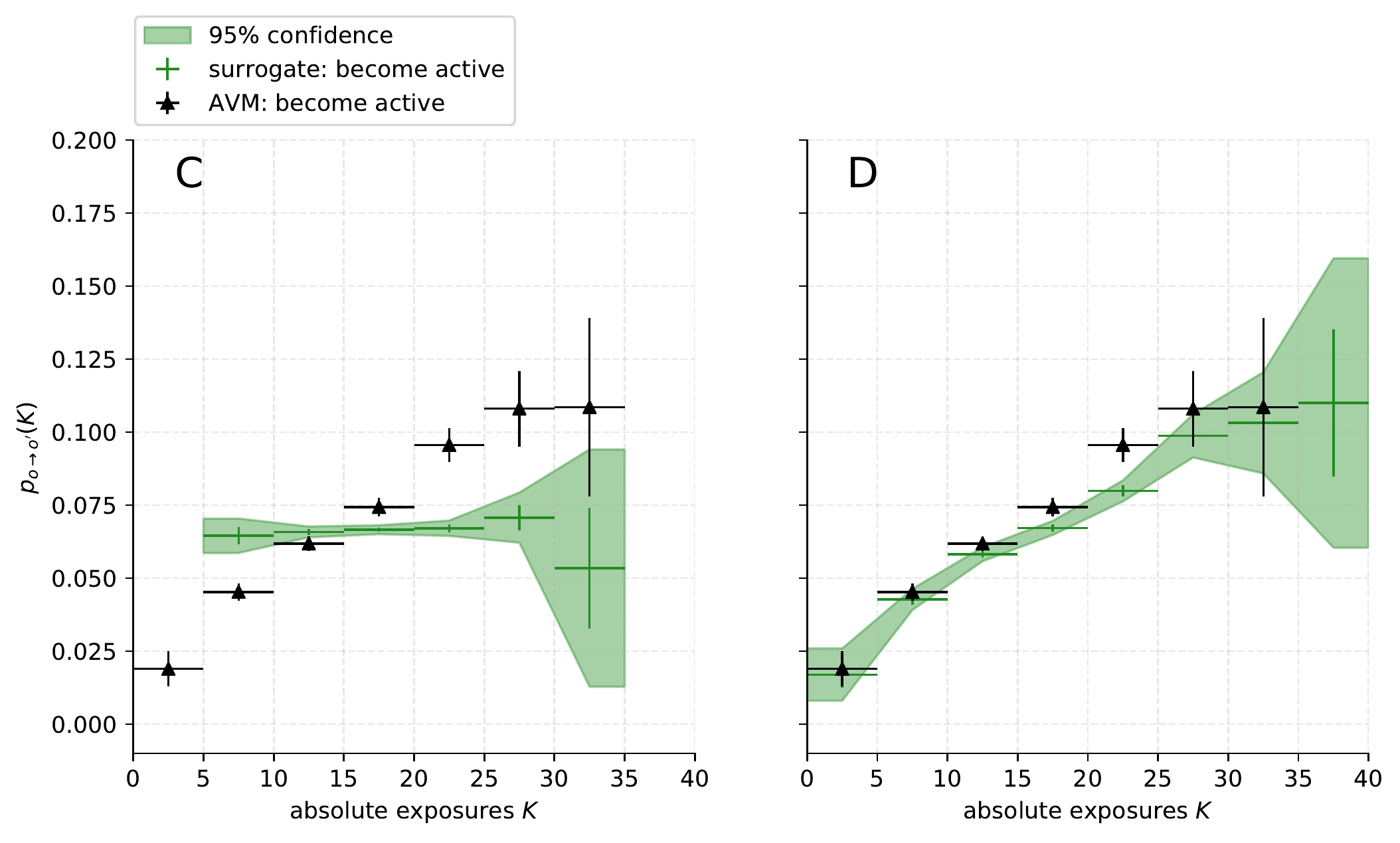}
	\caption{Comparison of DRFs computed on AVM data (black triangles) and surrogates of the node traits (green crosses), corresponding to the null hypothesis $\mathcal{H}_0^4$.
	(A) and (B) are for the $\varphi=0.0$ case, while (C) and (D) correspond to the $\varphi=0.6$ case.
	(B) and (D) display the same data as (A) and (C), respectively, but with y-axes independently shifted for AVM and surrogate data, to facilitate the direct comparison of the DRFs.
	%\jona{Error bars for the model (averaged over all model simulations???) and surrogate DRFs (averaged over all surrogate realizations) are based on a conservative estimate of the standard error of the switching probabilities $p(K)$ (Eq.~\ref{eq:error-bars}). Confidence bounds for surrogate DRFs are based on ... .}
	\niklas{Error bars are computed as described in Sect.~\ref{sec:dose_response_fct}.
	Confidence bounds for surrogate DRFs are the 95\,\% confidence interval of the distribution of $\p(K)$ over all surrogate realisations.}
	}
	\label{fig:edge_surrogates_avm}
\end{figure}

\paragraph{Fifth AVM Test. Hypothesis $\mathcal{H}_0^5$: $P(A, O_i(t))$:} Displayed in Fig.~\ref{fig:edge_surrogates_avm}A and C for $\varphi=0.0$ and $\varphi=0.6$, respectively.
As expected, after completely randomising the network, the surrogate model gives a nearly constant DRF. 
\jakob{The $\zeta$-score for the run with $\varphi = 0.0$ is $\zeta = 2.7$ and the score for $\varphi = 0.6$ is $\zeta=6.5$.}
The difference between surrogate and AVM DRFs is less significant for the $\varphi=0.0$ case than for the $\varphi=0.6$ case, which can be explained by the additional network processes at work in the latter case: the randomisation has a larger effect here.

\paragraph{Sixth AVM Test. Hypothesis $\mathcal{H}_0^6$: $P(k_i(t), O_i(t))$:} Displayed in Fig.~\ref{fig:edge_surrogates_avm}B and D for $\varphi=0.0$ and $\varphi=0.6$, respectively.
\jakob{The $\zeta$-score for the run with $\varphi = 0.0$ is $\zeta = 0.49$ and the score for $\varphi = 0.6$ is $\zeta=1.25$.}
The difference between the surrogate and original AVM DRFs is not nearly as big as in many of the other surrogate tests, pointing to an effect of homophily that is moderate at most. 
For the $\varphi=0.6$ case, hints for homophily effects can be observed, since the surrogate and original AVM curves are significantly separated here. For the $\varphi=0.0$ case, the curves are not significantly separated (see also Fig.~\ref{fig:surrogate_overview_avm}).
This is consistent with our expectations, since homophilic clustering through preferential attachment is present, but not dominant in the $\varphi=0.6$ model (see Fig.~\ref{fig:DRF_example})

\paragraph{}
Fig.~\ref{fig:surrogate_overview_avm} shows, analogously to Fig.~\ref{fig:surrogate_overview}, the significance of the deviations between surrogate and AVM DRFs
\jakob{for $\varphi = 0.0$}. The case $\varphi = 0.6$ was already presented in Fig.~\ref{fig:surrogate_overview}A.
For the $\varphi=0.0$ case (Fig.~\ref{fig:surrogate_overview_avm}), only 
%$\mathcal{H}_0^3$ and 
$\mathcal{H}_0^5$ cannot be rejected based on 
%Stouffer's z-score method.
\jakob{the $\zeta$ test statistic}.
For the $\varphi=0.6$ case (Fig.~\ref{fig:surrogate_overview}A), 
\jakob{none of the hypothesis tests can be rejected.}
%only $\mathcal{H}_0^3$ cannot be rejected.
\jakob{The difference in the rejection of $\mathcal{H}_0^3$ to the empirical case (Fig.~\ref{fig:surrogate_overview}B) 
\niklas{appears to show that our method can } 
detect contagion created by the social learning within the AVM.}
Moreover, the difference in the rejection of $\mathcal{H}_0^5$ 
\jakob{between the $\varphi = 0$ and the $\varphi = 0.6$ cases}
\niklas{suggests that our method can }
detect the small amount of homophily created by the adaptive rewiring.

\niklas{\section{Permutation space for \texorpdfstring{$\mathcal{H}_0^3$}{H03}}\label{apx:h03}}
For the surrogate method to work, the shuffling algorithms must provide sufficient randomisation, creating data sets with significant differences to the original data.
This is easily achieved for most of the proposed surrogate models.
However, the randomisation space for $\mathcal{H}^3_0$ is the most constrained.
Here, the number of possible permutations of the activity intervals is limited by the total number of activity level switches of each node.
In this section, we demonstrate that this randomisation space is sufficient for the method to function.

\begin{figure}[h]
	\centering
	\includegraphics[width=\Lscale\linewidth]{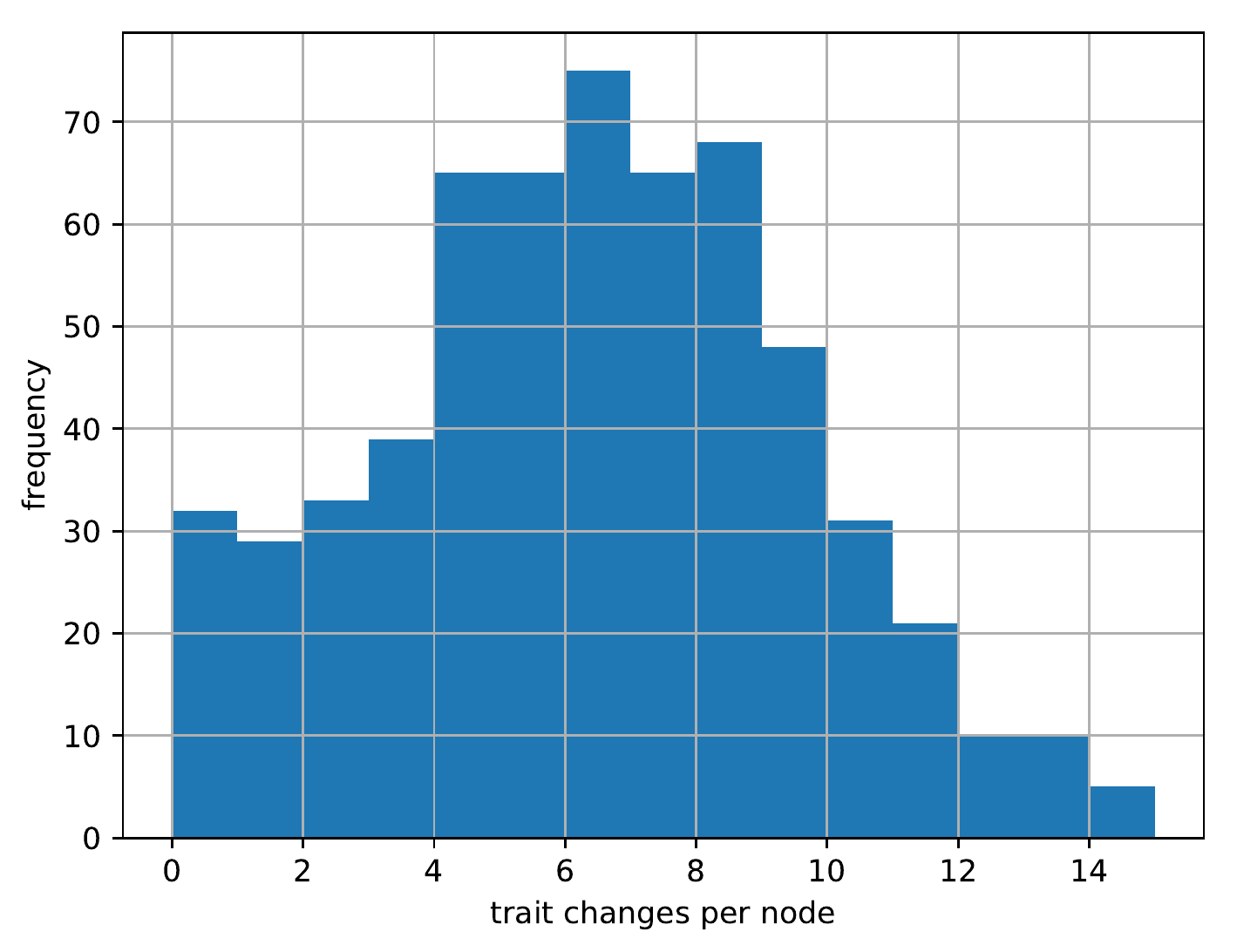}
	\caption{Distribution of the total number of total activity level changes per node during the studied time interval.
	}
	\label{fig:cns_node_trait_changes}
\end{figure}

Fig.~\ref{fig:cns_node_trait_changes} displays the distribution of total activity level (``trait'') changes per node in the studied time interval.
Nodes switch behaviour on average 5.94 times. 
Thus, on average, there are 3–4 active and 3–4 inactive intervals for each node. 
If a node has 3 active and 4 inactive intervals, the shuffling can produce 3!4! = 144 different surrogates. 
More than 43 percent of agents switch behaviour at least 7 times, thus having at least 4 active and 4 inactive intervals and hence at least 4!4! = 576 different surrogates for each of these nodes. 
From this, we conclude that there is sufficient randomisation in $\mathcal{H}^3_0$. 
This is supported by the validation of the methodology using synthetic AVM data, which shows a deviation between AVM and surrogate DRFs for $\mathcal{H}^3_0$ (see Fig.~\ref{fig:trait_surrogates_avm}C and F).

\FloatBarrier
\newpage

\section{List of considered fitness centers in Copenhagen}
\label{apx:locations}

\begin{table}[hbtp]
	\centering
	\begin{tabular}{lrr}
                                            Name &                                           Longitude [$^\circ$ E] & Latitude [$^\circ$ N] \\

					   \hline
                          Fresh Fitness Hvidovre &                      12.4691961 & 55.6415696\\
                                      Fitness.dk &                      12.5618214 & 55.6614733\\
                                       FitnessDK &                      12.5114098 & 55.6647699\\
                                   Fresh Fitness &                      12.5404751 & 55.6975516\\
                                           Fresh &                      12.4199488 & 55.6493081\\
                                   Fitness World &                      12.4418141 & 55.7231967\\
                          Fitness World Ballerup &                      12.3579672 & 55.7296181\\
                           Fitness World Brøndby &                      12.4383494 & 55.6673030\\
                        Fitness World Farum Park &                      12.3513120 & 55.8172970 \\
 Fitness World Frederiksberg Bernhard Bangs Alle &                      12.5104671 & 55.6844058\\
               Fitness World Frederiksberg Forum &                      12.5524718 & 55.6830906\\
     Fitness World Frederiksberg Peter Bangs Vej &                      12.5131680 & 55.6795400\\
                          Fitness World Gentofte &                      12.5378949 & 55.7386120\\
                          Fitness World Glostrup &                      12.4008395 & 55.6640800 \\
          Fitness World Greve Hundige Storcenter &                      12.3274148 & 55.5987709\\
                             Fitness World Greve &                      12.2984612 & 55.5905648\\
                            Fitness World Herlev &                      12.4160534 & 55.7253403\\
                             Fitness World Husum &                      12.4810239 & 55.7095419\\
      Fitness World København Baron Boltens Gård &                      12.5848511 & 55.6820125\\
            Fitness World København Ellebjergvej &                      12.5108247 & 55.6507568\\
          Fitness World København Emdrup Station &                      12.5409464 & 55.7218740\\
             Fitness World København Englandsvej &                      12.6043943 & 55.6569690\\
             Fitness World København Gasværksvej &                      12.5570237 & 55.6708078\\
                 Fitness World København Jagtvej &                      12.5509410 & 55.6964980 \\
               Fitness World København Lyngbyvej &                      12.5604444 & 55.7116463\\
                Fitness World København Lyongade &                      12.6099453 & 55.6613686\\
         Fitness World København Nordre Fasanvej &                      12.5364747 & 55.6985181\\
             Fitness World København Strandvejen &                      12.5777058 & 55.7219712\\
     Fitness World København Vester Farimagsgade &                      12.5623173 & 55.6782088\\
               Fitness World København Århusgade &                      12.5872772 & 55.7067752\\
                            Fitness World Lyngby &                      12.5039072 & 55.7688801\\
                             Fitness World Måløv &                      12.3187172 & 55.7485909\\
                            Fitness World Søborg &                      12.4932893 & 55.7395909\\
                          Fitness World Taastrup &                      12.3017208 & 55.6529634\\
                  Fitness World Valby Mosedalvej &                      12.5134815 & 55.6674858\\
                           Fitness World Værløse &                      12.3615021 & 55.7821745\\
                                       fitnessdk &                      12.4392816 & 55.7249089\\
	\end{tabular}
	\caption{List of the fitness centers in Copenhagen considered in this study, with their respective coordinates, as extracted from Open Street Maps \cite{OpenStreetMap}.}
	\label{tab:fitness_centers}
\end{table}

\end{document}